\documentclass[%
 reprint,
 superscriptaddress,
 preprintnumbers,
nofootinbib,
 amsmath,amssymb,
 aps,
]{revtex4-1}

\usepackage{graphicx}
 \usepackage{dcolumn}
\usepackage{bm}
\usepackage{color}
\usepackage[normalem]{ulem}

\usepackage{hyperref}

\newcommand{\fref}[1]{Fig.~\ref{fig:#1}} 
\newcommand{\eref}[1]{Eq.~\eqref{eq:#1}}

\newcommand{\aref}[1]{Appendix~\ref{app:#1}}

\newcommand{\cref}[1]{Chapter~\ref{ch:.#1}}
\newcommand{\tref}[1]{Table~\ref{tab:#1}}

\newcommand{\nn}{\nonumber \\}  
\newcommand{\nnl}{\nonumber \\}

\newcommand{\beq}{\begin{equation}} 
\newcommand{\eeq}{\end{equation}} 
\newcommand{\ba}{\begin{array}}  
\newcommand{\ea}{\end{array}} 
\newcommand{\bea}{\begin{eqnarray}}  
\newcommand{\eea}{\end{eqnarray} }  
\newcommand{\be}{\begin{eqnarray}}  
\newcommand{\ee}{\end{eqnarray} }  
\newcommand{\bal}{\begin{align}}
\newcommand{\eal}{\end{align}}   
\newcommand{\bi}{\begin{itemize}}  
\newcommand{\ei}{\end{itemize}}  
\newcommand{\ben}{\begin{enumerate}}  
\newcommand{\een}{\end{enumerate}}  
\newcommand{\bc}{\begin{center}}
\newcommand{\ec}{\end{center}} 
\newcommand{\bt}{\begin{table}}
\newcommand{\et}{\end{table}}  
\newcommand{\btb}{\begin{tabular}}
\newcommand{\etb}{\end{tabular}}  
\newcommand{\bvec}{\left ( \ba{c}}
\newcommand{\evec}{\ea \right )}

\newcommand{\cO}{{\mathcal O}} 
 
\newcommand{\cL}{{\mathcal L}}

\newcommand{\gev}{\mathrm{GeV}}

\def\hc{{\rm h.c.}} 

\newcommand{\eps}{\epsilon}

\newcommand{\eT}{\epsilon_T}
\newcommand{\eP}{\epsilon_P}
\newcommand{\eL}{\epsilon_L}
\newcommand{\eR}{\epsilon_R}

\begin{document}

\preprint{CERN-TH-2018-171}
\preprint{LA-UR-18-27408}
\preprint{LPT-Orsay-18-82}

\title{Hadronic tau decays as New Physics probes in the LHC era}

\author{Vincenzo Cirigliano}
\affiliation{Theoretical Division, Los Alamos National Laboratory, Los Alamos, NM 87545, USA}
\author{Adam Falkowski}
\affiliation{Laboratoire de Physique Th\'{e}orique (UMR8627), CNRS, Univ. Paris-Sud, Universit\'{e} Paris-Saclay, 91405 Orsay, France}
\author{Mart\'{i}n Gonz\'{a}lez-Alonso}%
\affiliation{Theoretical Physics Department, CERN, 1211 Geneva 23, Switzerland}
\author{Antonio Rodr\'iguez-S\'anchez}
\affiliation{Departament de F\'isica Te\`orica, IFIC, Universitat de Val\`encia - CSIC, Apt.  Correus 22085, E-46071 Val\`encia, Spain}
\affiliation{Department of Astronomy and Theoretical Physics, Lund University, S\"olvegatan 14A, SE 223-62 Lund, Sweden}

\begin{abstract}

We analyze the sensitivity of hadronic $\tau$ decays to non-standard 
interactions within the model-independent 
framework of the Standard Model Effective Field Theory (SMEFT). 
Both exclusive and inclusive decays are studied, using the latest lattice data and  QCD dispersion relations. 
We show that there are enough theoretically clean channels to disentangle all
 the effective couplings contributing to 
these decays, with the $\tau\to\pi\pi\nu_\tau$ channel representing an unexpected powerful New Physics probe. 
We find that the ratios of  non-standard couplings to the Fermi constant are bound at the sub-percent level. 
These  bounds are  complementary to the ones from electroweak precision observables  and  $pp\to\tau\nu_\tau$ measurements at the LHC.
 The combination of $\tau$ decay and LHC data puts tighter constraints on lepton universality 
 violation in the gauge boson-lepton vertex corrections. 
 
\end{abstract}

\maketitle

Hadronic tau decays have been extensively used in the last decades to learn about fundamental physics~\cite{Pich:2013lsa,Schael:2005am}. The inclusive decays are used to accurately extract fundamental Standard Model (SM) parameters such as the strong coupling constant~\cite{Braaten:1991qm,Boito:2014sta,Pich:2016bdg}, the strange quark mass or the $V_{us}$ entry of the Cabibbo-Kobayashi-Maskawa (CKM) matrix~\cite{Gamiz:2002nu,Gamiz:2004ar}. They also represent a valuable QCD laboratory, where chiral parameters or properties of the QCD vacuum can be extracted with high precision in a model-independent fashion through dispersion relations~\cite{Boito:2015fra,Rodriguez-Sanchez:2016jvw}. 
On the other hand, exclusive hadronic tau decays are much harder to predict within QCD with high accuracy, and thus, they are useful for learning about hadronic physics. The only exceptions are the two-body decays $\tau\to\pi \nu_\tau,K\nu_\tau$, thanks to the precise lattice calculations of the pion and kaon decay constants~\cite{Aoki:2016frl}. 

The agreement between the above-mentioned determinations of SM and QCD parameters with determinations using other processes represents a non-trivial achievement, which is only possible thanks to the impressive effort carried out in several fronts: experimental, lattice and analytical QCD methods. Needless to say, this agreement is easily spoiled if non-standard effects are present. However, the use of hadronic tau decays as New Physics (NP) probes has been marginal so far (see e.g. Refs.~\cite{Bernard:2007cf,Garces:2017jpz}), with the exception once again of the simple $\tau\to\pi \nu_\tau,K\nu_\tau$ channels. The goal of this letter is to amend this situation presenting an unprecedented comprehensive analysis of the NP reach of $CP$-conserving hadronic tau decays.

For  sake of definiteness, we focus on the non-strange decays, which are governed by the following low-energy effective Lagrangian~\cite{Cirigliano:2009wk}~\footnote{We have not included right-handed (and wrong-flavor) neutrino fields~\cite{Cirigliano:2012ab}, which in any case do not interfere with the SM amplitude and thus contribute at $\mathcal{O}(\epsilon_i^2)$ to the observables.
}
\bea
{\cal L}_{\rm eff} 
&=& - \frac{G_F V_{ud}}{\sqrt{2}}  \Bigg[
\Big(1 + \epsilon_L^{ \tau}  \Big) \bar{\tau}  \gamma_\mu  (1 - \gamma_5)   \nu_{\tau} \cdot \bar{u}   \gamma^\mu (1 - \gamma_5 ) d
\nn
&&+  \eR^{\tau}  \   \bar{\tau} \gamma_\mu (1 - \gamma_5)  \nu_\tau    \cdot \bar{u} \gamma^\mu (1 + \gamma_5) d
\nonumber\\
&&+~ \bar{\tau}  (1 - \gamma_5) \nu_{\tau} \cdot \bar{u}  \Big[  \epsilon_S^{\tau}  -   \eP^{\tau} \gamma_5 \Big]  d
\nn
&&+ \epsilon_T^{\tau}    \,   \bar{\tau}   \sigma_{\mu \nu} (1 - \gamma_5) \nu_{\tau}    \cdot  \bar{u}   \sigma^{\mu \nu} (1 - \gamma_5) d
\Bigg]+{\rm h.c.}, 
\label{eq:leff1} 
\eea
where we use $\sigma^{\mu \nu} = i\,[\gamma^\mu, \gamma^\nu]/2$ and $G_F$ is the Fermi constant. 
The only assumptions are Lorentz and $U(1)_{\rm em}\times SU(3)_C$ invariance,   and the absence of light nonstandard particles. In practice we also assume that the subleading derivative terms in the EFT expansion (suppressed by $m_\tau/m_W$) are indeed negligible. The Wilson coefficients $\epsilon_i$ parametrize non-standard contributions, and they vanish in the SM leaving the $V-A$ structure generated by the exchange of a $W$ boson. 
 The nonstandard coefficients $\epsilon_i$ can be complex, but the sensitivity of the observables considered in this work to the imaginary parts of the coefficients is very small. Thus, the results hereafter implicitly refer to the real parts of $\epsilon_i$.

Through a combination of inclusive and exclusive  $\tau$ decays, 
we are able to constrain all the Wilson coefficients  in \eref{leff1} -- this is the main result of this paper. 
In  the $\overline{\rm MS}$ scheme  at   scale $\mu=2$ GeV we  
find the following central values and $1\sigma$ uncertainties:
\bea
\label{eq:epsconstraints}
\left(
\begin{array}{c}
\eL^{\tau}\!-\!\eL^{e}\!+\!\eR^{\tau}\!-\!\eR^{e} \\
\epsilon_R^{ \tau} \\
\epsilon_S^{ \tau} \\
\epsilon_P^{ \tau} \\
\epsilon_T^{ \tau} \\
\end{array}
\right)
=
\left(
\begin{array}{c}
  1.0 \pm 1.1  \\
 0.2 \pm 1.3 \\
 -0.6 \pm 1.5  \\
 0.5 \pm 1.2 \\
 -0.04 \pm 0.46 \\
\end{array}
\right)\cdot 10^{-2},
\eea
where $\epsilon_{L,R}^e$ parametrize electron couplings to the first generation quarks and are defined in analogy to their tau counterparts. 
They affect the $G_F V_{ud}$ value obtained in nuclear $\beta$ decays~\cite{Gonzalez-Alonso:2018omy}, which is needed in the analysis of hadronic tau decays. 
The correlation matrix associated to~\eref{epsconstraints} is
\beq 
\label{eq:correlation}
\rho  = 
\left ( \ba{cccc}  
0.88  & 0 & -0.57 & -0.94 \\ 
  & 0 & -0.86 & -0.94 \\
  &  & 0 & 0 \\
  &  &  & 0.66 \\
\ea
\right )
.
\eeq
Below we summarize how Eqs.~(\ref{eq:epsconstraints})-(\ref{eq:correlation}) were derived. 

\section{Exclusive decays} 
The $\tau\to\pi \nu_\tau$ channel~\cite{Patrignani:2016xqp} gives the following 68\% CL constraint:
\bea
\label{eq:pi}
\eL^{\tau} - \eL^{e} - \eR^{\tau} - \eR^{e} - \frac{B_0}{m_\tau} \eP^{\tau} = (-1.5 \pm 6.7)\cdot 10^{-3}~,
\eea
where $B_0 = m_\pi^2/(m_u + m_d)$. We included the SM radiative corrections~\cite{Decker:1994ea,Cirigliano:2007xi,Rosner:2015wva} and the latest lattice average for the pion decay constant, $f_{\pi^\pm} = 130.2(8)$ MeV ($N_f=2+1$)~\cite{FLAG2017}, from Refs.~\cite{Blum:2014tka,Bazavov:2010hj,Follana:2007uv}. 
We stress that the lattice determinations of $f_{\pi^\pm}$ are a crucial input to search for NP in this channel, and despite its impressive precision, it represents the dominant source of error in~\eref{pi}, 
followed by the experimental error (2.4 times smaller), and the radiative corrections uncertainty. Because of this, significant improvement in the bound above can be expected in the near future. 
Alternatively, as is often seen in the literature~\cite{Pich:2013lsa}, one can obtain tighter constraints on the effective theory parameters by considering ``theoretically clean'' ratios of observables where the $f_\pi$ dependence cancels out. For example, from the ratio $\Gamma(\tau\to\pi \nu)/\Gamma(\pi\to \mu \nu)$ one can deduce 
\beq
\label{eq:ratio}
\eps_L^\tau - \eps_L^\mu   - \eps_R^\tau +  \eps_R^\mu -   {B_0 \over m_\tau} \eps_P^\tau +  {B_0   \over m_\mu} \eps_P^\mu  =  (-3.8 \pm 2.7)\cdot 10^{-3} . 
\eeq 
This and similar constraints are {\em not} included in \eref{epsconstraints}, which only summarizes the input from hadronic tau decays without using any meson decay observables. 
Instead, we later combine \eref{epsconstraints} with the results of Ref.~\cite{Gonzalez-Alonso:2016etj}, which derived a likelihood for the effective theory parameters based on a global analysis of pion and kaon decays. 
The combination effectively includes constraints from $\Gamma(\tau\to\pi \nu)/\Gamma(\pi\to \ell \nu)$, with correlations due to the common $f_\pi$ uncertainty taken into account.

The $\tau\to\pi\pi\nu_\tau$ channel, which is sensitive to vector and tensor interactions, is much more complicated to predict within QCD in a model-independent way. However, a stringent constraint can be obtained through the comparison of the spectral functions extracted from $\tau\to\pi\pi\nu_\tau$ and its isospin-rotated process $e^+e^-\to\pi^+\pi^-$, after the proper inclusion of isospin-symmetry-breaking corrections. The crucial point here is that heavy NP effects (associated with the scale $\Lambda$) 
can be entirely neglected  in $e^+e^-\to\pi^+\pi^-$  at  energy $\sqrt{s} \ll \Lambda$ due to  the  electromagnetic nature of this process. 
We can benefit from past studies that exploited this isospin relation to extract from data the $\pi\pi$ component of the lowest-order hadronic vacuum polarization contribution to the muon $g-2$, usually denoted by $a_\mu^{\rm{had,LO}}[\pi\pi]$, through a dispersion integral. Such an approach implicitly assumes the absence of NP effects, which however may contribute to the extraction from tau data. In this way we find a sub-percent level sensitivity to NP effects:
\bea
\label{eq:pipi}
\frac{a_\mu^{\tau}\!-\!a_\mu^{ee}}{2\,a_\mu^{ee}} \!=\! \eL^{\tau}\!-\!\eL^{e}\!+\!\eR^{\tau}\!-\!\eR^{e} +\!1.7\,\eT^{\tau} \!=\! (8.9 \! \pm \! 4.4) \! \cdot \!10^{-3} \! , \quad 
\eea
where $a_\mu^{\tau} = (516.2 \pm 3.6)\times10^{-10}$
~\cite{Davier:2013sfa} and $a_\mu^{ee} = (507.14\pm 2.58)\times10^{-10}$
~\cite{Davier:2017zfy} are the values of $a_\mu^{\rm{had,LO}}[\pi\pi]$ extracted from $\tau$ and $e^+e^-$ data.
The $\sim\!\!2\sigma$ tension with the SM reflects the well-known disagreement between both datasets~\cite{Davier:2017zfy}.\footnote{Recently Ref.~\cite{Keshavarzi:2018mgv} found $a_\mu^{ee} = 503.74\pm 1.96$ using similar data but a different averaging method than Ref.~\cite{Davier:2013sfa}. While the change in the uncertainty is not significant, 
this increases to $3\sigma$ the tension between $a_\mu^{\tau}$ and $a_\mu^{ee}$.}
In order to estimate the factor multiplying  $\epsilon_{T}$  in~\eref{pipi}, 
we have  (i) assumed  that the proportionality of 
the tensor and vector form factors, which is exact in the  elastic region~\cite{Cirigliano:2017tqn,Miranda:2018cpf},   
holds in the dominant $\rho$ resonance region (as is the case within the resonance chiral theory 
framework~\cite{Ecker:1988te});
(ii) used the lattice QCD result of Ref. \cite{Baum:2011rm}  
for the $\pi\pi$ tensor form factor at zero momentum transfer (see also Refs.~\cite{Mateu:2007tr,Cata:2008zc}).
Inelastic effects impact the estimate 1.7  at the  $<10$~\% level. 
This  small uncertainty can be traced back to the fact that the coefficient 1.7  arises from the ratio of  two integrals over the $\pi \pi$ invariant mass, each involving the product of a rapidly decreasing weight function (which de-emphasizes the 
inelastic region) and appropriate form factors (whose uncertainty tends to cancel in the ratio). Details will be provided in Ref.~\cite{companion}.

The constraint above can be strengthened by directly looking at the $s$-dependence of the spectral functions (instead of the $a_\mu$ integral), which would also allow us to disentangle the vector and tensor interactions.\footnote{Useful angular and kinematic distributions including NP effects were recently derived in Ref.~\cite{Miranda:2018cpf}. 
}  
Moreover, the $a_\mu^{\tau,ee}$ uncertainties include a scaling factor due to internal inconsistencies of the various datasets~\cite{Davier:2017zfy}, which will hopefully decrease in the future. In fact, new analyses of the $\pi\pi$ channel are expected from CMD3, BABAR, and possibly Belle-2~\cite{Davier:2017zfy,Keshavarzi:2018mgv}. 
Last, $a_\mu^{\tau}$ will benefit from the ongoing calculations of isospin-breaking effects in the lattice~\cite{Bruno:2018ono}. 
All in all, we can expect a significant improvement in precision with respect to the result in~\eref{pipi} in the near future.

As recently pointed out in Ref.~\cite{Garces:2017jpz}, a third exclusive channel that can provide useful information is $\tau\to\eta\pi\nu_\tau$, since the non-standard scalar contribution is enhanced with respect to the (very suppressed) SM one. Because of this, one can obtain a nontrivial constraint on $\epsilon_S^\tau$ even though both SM and NP contributions are hard to predict with high accuracy. Using the latest experimental results for the branching ratio~\cite{delAmoSanchez:2010pc,Patrignani:2016xqp} and a very conservative estimate for the theory errors~\cite{Garces:2017jpz,Escribano:2016ntp,Roig:private} we find 
\bea
\eps^\tau_S = (-6\pm 15)\times 10^{-3}~, 
\label{eq:eta}
\eea
which will significantly improve if theory or experimental uncertainties can be reduced. The latter will certainly happen with the arrival of Belle-II, which is actually expected to provide the first measurement of the SM contribution to this channel~\cite{Kou:2018nap} (see also Ref.~\cite{Hayasaka:2009zz} for Belle results). This is the only probe in this work with a significant sensitivity (via $\cO(\eps_S^2)$ effects) to the imaginary part of $\epsilon_i$ coefficients. Including the latter does not affect the bound in~\eref{eta} though.

\section{Inclusive decays}
Summing over certain sets of decay channels one obtains the so-called inclusive vector (axial) spectral functions $\rho_{V(A)}$~\cite{Pich:2013lsa,Schael:2005am}. In the SM they are proportional to the imaginary parts of the associated $VV$ ($AA$) two-point correlation functions, $\Pi_{VV(AA)}(s)$, but these relations are modified by NP effects~\cite{Gonzalez-Alonso:2010vnm,Rodriguez-Sanchez:2018}. Thus, one could directly use the latest measurements of these spectral functions to constrain such effects if we had a precise theoretical knowledge of their QCD prediction.  
However, perturbative QCD is known not to be valid at $\sqrt{s} < 1\, \gev$, 
especially in the Minkowskian axis, where the spectral function lies. Nevertheless, one can make precise theoretical predictions for integrated quantities exploiting the well-known analiticity properties of QCD correlators~\cite{Braaten:1991qm}. Here we extend the traditional approach to also include NP effects, finding~\cite{Gonzalez-Alonso:2010vnm,Rodriguez-Sanchez:2018}
\bea
&&\int^{s_{0}}_{4m_\pi^2}\frac{ds}{s_{0}} \omega\left(\frac{s}{s_{0}}\right)\,\rho_{V\pm A}^{\rm{exp}}(s) 
\approx 
\left(1 + 2 \epsilon_V\right) X_{VV} \nonumber\\
&& \pm \left(1 + 2 \epsilon_A\right) \left( X_{AA} - \frac{f_\pi^2}{s_0}\omega\left(\frac{m_\pi^2}{s_{0}}\right) \right) + \eT^\tau\,X_{VT}~,
\label{eq:mastereq}
\eea
where $\omega(x)$ is a generic analytic function and $\rho_{V\pm A}^{\rm{exp}}(s)$ is the sum/difference of the vector and axial spectral functions, extracted experimentally under SM assumptions~\cite{Schael:2005am,Davier:2013sfa}. We also introduced the NP couplings $\epsilon_{V/A}\equiv \epsilon_{L\pm R}^{\tau}-\epsilon_{L+R}^{e}$, where $\epsilon^{\ell}_{L\pm R}\equiv \epsilon^{\ell}_L \pm \epsilon^{\ell}_R$. 
Last, $X_{ij}$ are QCD objects that can be 
calculated via the Operator Product Expansion, as discussed in~\aref{inclusive}. 
\eref{mastereq} shows how the agreement between precise SM predictions (RHS) and experimental results (LHS) for inclusive decays can be translated into strong NP constraints. 

In the $V+A$ channel, we find two clean NP constraints using $\omega(x)=(1-x)^{2}(1+2x)$, which gives the total nonstrange BR, and with $\omega(x)=1$. They provide respectively 
\begin{align}
\epsilon_{L+R}^{\tau}-\epsilon_{L+R}^{e}-0.78\epsilon_{R}^{\tau}+1.71\epsilon_{T}^{\tau}&=(4\!\pm\!16) \cdot 10^{-3}~,\label{eq:inclusive1}\\
\epsilon_{L+R}^{\tau}-\epsilon_{L+R}^{e}-0.89\epsilon_{R}^{\tau}+0.90\epsilon_{T}^{\tau}&\!=\!(8.5\! \pm\! 8.5)\! \cdot\! 10^{-3}.\label{eq:inclusive2}
\end{align}
The uncertainty in~\eref{inclusive1} comes mainly from the non-perturbative corrections, whereas that of~\eref{inclusive2} is dominated by experimental and Duality Violations (DV) uncertainties
(see~\aref{inclusive}).

In the $V-A$ channel, where the perturbative contribution is absent, two strong constraints can be obtained using $\omega(x)=1-x$ and  $\omega(x)=(1-x)^{2}$:
\begin{align}
\epsilon_{L+R}^{\tau}-\epsilon_{L+R}^{e}+3.1\epsilon_{R}^{\tau}+8.1\epsilon^{\tau}_{T}&= (5.0 \pm 50) \cdot 10^{-3} \, ,\\
\epsilon_{L+R}^{\tau}-\epsilon_{L+R}^{e}+1.9\epsilon_{R}^{\tau}+8.0\epsilon^{\tau}_{T}&= (10 \pm 10) \cdot 10^{-3} \, .\label{eq:inclusive4}
\end{align}
DV dominate uncertainties for the first constraint, while experimental and $f_{\pi}$ uncertainties dominate the latter one. This constraint could be improved with more precise data and $f_{\pi}$ calculations, but at some point DV, much more difficult to control, would become the leading uncertainty. 
The non-neglibible correlations between the various NP constraints derived above (due to $f_{\pi}$ and experimental correlations) have been taken into account in~\eref{epsconstraints}.

The weight functions chosen above are motivated by simplicity (low-degree polynomials), small non-perturbative  corrections, and different enough behavior so that their correlations can be taken into account.
\\

\section{Electroweak Precision Data}

If NP is coming from dynamics at $\Lambda \gg m_Z$ and electroweak symmetry breaking is linearly realized, then the relevant effective theory at $E \gtrsim m_Z$ is the so-called SMEFT, which has the same local symmetry and field content as the SM, but also contains higher-dimensional operators encoding NP effects ~\cite{Buchmuller:1985jz,Cirigliano:2009wk,Grzadkowski:2010es}.  
The SMEFT framework allows one to combine in a model-independent way constraints from low-energy measurements with those from Electroweak Precision Observables (EWPO) and LHC searches.
Moreover, once the SMEFT is matched to concrete UV models at the scale $\Lambda$, one can efficiently constrain masses and couplings of NP particles. 
The dictionary between low-energy parameters in \eref{leff1} and Wilson coefficients in the Higgs basis~\cite{deFlorian:2016spz,Falkowski:2017pss} is
\bea
\label{eq:epstosmeft}
\eL^{ \tau} - \eL^{ e}   &= &   \delta g_L^{W \tau}  -  \delta g_L^{W e} - [c^{(3)}_{\ell q}]_{\tau\tau11} + [c^{(3)}_{\ell q}]_{ee11}~,
\nnl 
\epsilon_R^{\tau} &= &  \delta g_R^{W q_1} , 
\nnl 
\epsilon_{S,P}^{ \tau} &=& 
- \frac{1}{2} [c_{lequ}\pm c_{ledq}]^*_{\tau\tau11}~,
\nonumber\\
\epsilon_T^{ \tau} &=& 
- \frac{1}{2} [c^{(3)}_{lequ}]^*_{\tau\tau11}~,
\eea 
where we approximate $V_{\rm CKM}\approx {\bf 1}$ in these $\cO(\Lambda^{-2})$ terms. 
The coefficients $\delta g_{L/R}^{Wf}$ are corrections to the SM $Wff'$ vertex and $c_i/v^2$ parametrize 4-fermion interactions with different helicity structures ($v\approx 246$ GeV); 
see~\aref{SMEFT} for their precise definitions.

Note that $\epsilon_R^{\ell}$ is lepton-universal in the SMEFT, up to dim-8 corrections~\cite{Bernard:2006gy,Cirigliano:2009wk}.  
We perform this matching at $\mu=M_Z$, after taking into account the QED and QCD running of the low-energy coefficients $\epsilon_i$ up to the electroweak (EW) scale~\cite{Gonzalez-Alonso:2017iyc}. Electroweak and QCD running to/from 1 TeV is also carried out in the comparison with LHC bounds below.
The running is numerically important for (pseudo-)scalar and tensor operators, influencing the confidence intervals at an $\cO(100\%)$ level and introducing mixing between the corresponding Wilson coefficients.

Our results are particularly relevant for constraining lepton flavor universality (LFU) violation, which can be done through an SMEFT analysis with all dimension-6 operators present simultaneously. As a matter of fact, Ref.~\cite{Falkowski:2017pss} carried out a flavor-general SMEFT fit to a long list of precision observables, which did not include however any observable sensitive to $qq\tau\tau$ interactions. 
As a result, no bound was obtained on the four-fermion Wilson coefficients, $[c_i]_{\tau\tau11}$. 
From \eref{epstosmeft}, given that $[c^{(3)}_{\ell q}]_{ee11}$ and the vertex corrections $\delta g$ are independently constrained, hadronic tau decays imply novel limits on these coefficients. 
We find 
\beq 
\label{eq:smeftbounds}
\left [ \ba{c} c^{(3)}_{lq} \\ c_{lequ} \\ c_{ledq} \\ c^{(3)}_{lequ} \ea \right ]_{\tau\tau11} 
\!\!\!\!\!\!= 
\bvec\! 0.012(29) \\  \!-0.002(11) \\  \!0.009(11) \\  \!-0.0036(93) \!\!\evec ~
\footnotesize{
\rho \!=\! \left (\! \ba{ccc} 
\!.09 & \!-.09 & \!.02 \\ 
 & \!.37 & \!.29 \\
 &  & \!-.28 \\
\ea \!\!\right )}~, 
\eeq 
after marginalizing over the remaining SMEFT parameters.   
These are not only very strong but also unique low-energy bounds.
On the other hand Ref.~\cite{Falkowski:2017pss} did access the right-handed vertex correction: 
$\delta g_R^{W q_1} = -(1.3\pm1.7)\times 10^{-2}$, 
from neutron beta decay~\cite{Gonzalez-Alonso:2016etj,Alioli:2017ces}. 
Including hadronic tau decays in the global fit improves this significantly: $\delta g_R^{W q_1} = -(0.4 \pm 1.0)\times 10^{-2}$.\footnote{A 50\% stronger (weaker) bound on $\delta g_R^{W q_1}$ is obtained using the recent lattice determination of the axial charge in Ref.~\cite{Chang:2018uxx} (Ref.~\cite{Gupta:2018qil}).} 
The fact that $\epsilon_R^\tau$, probed by tau decays, and $\epsilon_R^e$, probed by beta decays, are connected to one and the same  parameter $\delta g_R^{W q_1}$ is a prediction of the SMEFT, and would not be true in a more general setting where EW symmetry is realized non-linearly.  
Thus, comparison between phenomenological determinations of  $\epsilon_R^\tau$ and  $\epsilon_R^e$ (both consistent with zero currently) provides a test of that SMEFT assumption.


%
\begin{table}[h]
\centering
\begin{tabular}{ | c | c | c  | c|}
 \hline
 Coefficient			&	ATLAS $\tau \nu$		& $\tau$ decays  & $\tau$ and $\pi$ decays  \\ \hline			
$[c^{(3)}_{\ell q}]_{\tau\tau11}$ 	&	 $[0.0,1.6]$      &$ [-12.6,0.2]$  &$ [-7.6,2.1]$  \\
$[c_{\ell e q u }]_{\tau\tau11} $	&	$[-5.6,5.6]$	&	$[-8.4,4.1]$ &	$[-5.6,2.3]$	 \\
 $ [c_{\ell e d q}]_{\tau\tau11} $	&	$[-5.6,5.6]$	&	$[-3.5,9.0]$ &	$[-2.1,5.8]$ \\
$[c^{(3)}_{\ell e q u}]_{\tau\tau11} $&	$[-3.3,3.3]$	&  $[-10.4,-0.2]$ &  $[-8.6,0.7]$  \\ \hline
\end{tabular}
\caption{95\% CL intervals (in $10^{-3}$ units) at $\mu = 1$~TeV, assuming one Wilson coefficient is present at a time. 
The third column uses~\eref{epsconstraints}, whereas the fourth one also includes clean LFU ratios such as $\Gamma(\tau\to\pi \nu)/\Gamma(\pi\to \mu \nu)$. 
}
\label{tab:LHC}
\end{table}
 
\section{LHC bounds} 
It is instructive to compare the NP sensitivity of hadronic tau decays to that of the LHC. 
While the experimental precision is typically inferior for the LHC, it probes much higher energies and may offer a better reach for the Wilson coefficients whose contribution to observables is enhanced by $E^2/v^2$.    
We focus on the high-energy tail of the $\tau \nu$ production.  
This process is sensitive to the 4-fermion coefficients 
$[c^{(3)}_{\ell q},c_{\ell e q u },c_{\ell e d q}, c^{(3)}_{\ell e q u}]_{\tau\tau11}$, which also affect tau decays. 
Other Wilson coefficients in~\eref{epstosmeft} do not introduce energy-enhanced corrections to the $\tau \nu$ production, and can be safely neglected in this analysis.\footnote{Other energy-enhanced operators do not interfere with the SM. Thus, their inclusion would not change our analysis.} 
 
In~\tref{LHC} we show our results based on a recast of the transverse mass $m_T$ distribution of $\tau \nu$ events in $\sqrt{s}=13$~TeV LHC collisions recently measured by ATLAS \cite{Aaboud:2018vgh}.  
We estimated the impact of the Wilson coefficients on the $d \sigma(p p \to \tau \nu) / d m_T$ cross section using the {\tt Madgraph}\cite{Alwall:2014hca}/{\tt Pythia~8}~\cite{Sjostrand:2014zea}/{\tt Delphes} \cite{deFavereau:2013fsa} simulation chain. 
We assign 30\% systematic uncertainty to that estimate, which roughly corresponds to the size of the NLO QCD corrections to the NP terms (not taken into account in our simulations)~\cite{Alioli:2018ljm}.   
The SM predictions are taken from  \cite{Aaboud:2018vgh}, and their quoted uncertainties in each bin are treated as independent nuisance parameters. 
We find that for the chirality-violating operators the LHC bounds are comparable to those from hadronic tau decays. 
On the other hand, for the chirality-conserving coefficient $[c^{(3)}_{\ell q}]_{\tau\tau11}$ the LHC bounds are an order of magnitude stronger thanks to the fact that the corresponding operator interferes with the SM $q \bar q' \to \tau \nu$ amplitude.  
Let us stress that SMEFT analyses of high-$p_T$ data require additional assumptions though, such as heavier NP scales and suppressed dimension-8 operator contributions. 
Last, we observe an $\cO(2)\sigma$ preference for a non-zero value of $[c^{(3)}_{\ell q}]_{\tau\tau11}$ due to a small excess over the SM prediction observed by ATLAS in several bins of the $m_T$ distribution.

\begin{figure}[!htb]
\begin{center}
\includegraphics[width=0.95\columnwidth]{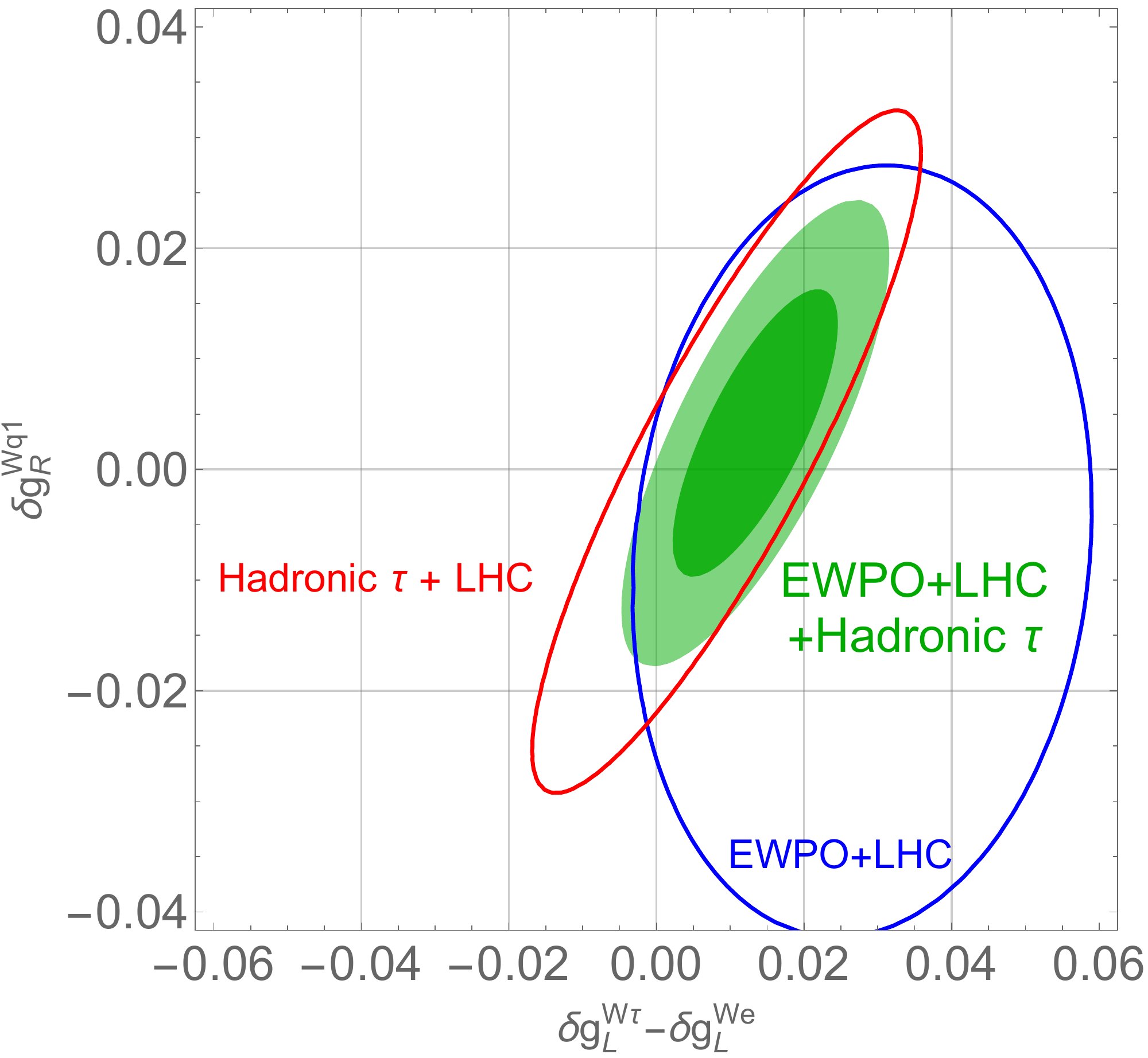}
\caption{68\% and 95\% CL bounds on the vertex corrections $\delta g_L^{W \tau}  -  \delta g_L^{W e}$ and $ \delta g_R^{W q_1}$ (green) after using the LHC input to constrain $[c^{(3)}_{\ell q}]_{\tau \tau 11}$ and  $[c^{(3)}_{\ell q}]_{ee11}$.  We also show separate 95\% CL contours from hadronic $\tau$ decays (red) and from previous EWPO~\cite{Falkowski:2017pss} (blue), which are correlated, {\it cf.} discussion below~\eref{ratio}. 
\label{fig:vertex}
}
\end{center}
\end{figure}

The LHC and $\tau$ decay inputs together allow us to sharpen the constraints on LFU of gauge interactions.
From \tref{LHC},  $[c^{(3)}_{\ell q}]_{\tau \tau 11}$ is constrained by the LHC at an $\cO(10^{-3})$ level, 
and similar conclusions can be drawn with regard to $[c^{(3)}_{\ell q}]_{ee11}$ \cite{Falkowski:2017pss,Greljo:2017vvb}. 
Then hadronic tau decays  effectively become a new probe of the vertex corrections: $\eL^{ \tau} - \eL^{ e}\approx \delta g_L^{W \tau}  -  \delta g_L^{W e}$ and $\epsilon_R^{ \tau}=\delta g_R^{W q_1}$, complementing the information from previous low-energy EWPO~\cite{Falkowski:2017pss}. 
The interplay between the two is shown in~\fref{vertex}. 
The input from hadronic tau decays leads to the model independent constraint on LFU of $W$ boson interactions:  $\delta g_L^{W \tau}  -  \delta g_L^{W e} = 0.0134(74)$, 
which becomes more than two times stronger in the simpler scenario where $\delta g_L^{W \tau}  -  \delta g_L^{W e}$ is the only deformation of the SM.

\section{Conclusions} 
We have shown in this work that hadronic $\tau$ decays represent competitive NP probes, thanks to the very precise measurements and SM calculations. This is a change of perspective with respect to the usual approach, which considers these decays as a QCD laboratory where one can learn about hadronic physics or extract fundamental parameters such as the strong coupling constant. From this new perspective, the agreement between such determinations~\cite{Braaten:1991qm,Boito:2014sta,Pich:2016bdg} and that of Ref.~\cite{Aoki:2016frl} in the lattice is recast as a stringent NP bound. Our results are summarized in~\eref{epsconstraints} and can be easily applied to constrain a large class of NP models with the new particles heavier than $m_\tau$.
Hadronic $\tau$ decays probe new particles with up to $\cO(10)$~TeV masses (assuming order one coupling to the SM) or even $\cO(100)$~TeV masses,  for strongly coupled  scenarios. 
They can be readily combined with other EWPO within the SMEFT framework to constrain NP heavier than $m_Z$.
Including this new input in the global fit leads to four novel constraints in \eref{smeftbounds},  which are the first model-independent bounds on the corresponding $\tau \tau q q$ operators. 
Moreover, it leads to tighter bounds on the $W$ boson coupling to right-handed quarks.    
Hadronic $\tau$ decays represent a novel sensitive probe of LFU violation ($\tau$ vs. $e$), which competes with and greatly complements EWPO and LHC data. 
This is illustrated in~\fref{vertex} and~\tref{LHC} for vertex corrections and contact interactions respectively.
Thus, our constraints can be useful in relation with the current hints of LFU violation in certain $B$ meson decays~\cite{Lees:2013uzd,Huschle:2015rga,Sato:2016svk,Aaij:2014ora,Aaij:2017vbb}, or the old tension in $W$ decays~\cite{Filipuzzi:2012mg,Patrignani:2016xqp}.
For instance, our model-independent $\cO(1)\%$ constraints in~\eref{epsconstraints} imply that the hints for $\cO(10)\%$ LFU violation observed in $B \to D^{*} \tau \nu$ decays \cite{Lees:2013uzd,Huschle:2015rga,Sato:2016svk} cannot be explained by NP effects in the hadronic decay of the $\tau$ lepton, but must necessarily involve (as is the case in most models) non-standard LFU-violating interactions involving the bottom quark.    

The discovery potential of these processes in the future is very promising since the constraints derived in this work are expected to improve with the arrival of new data (e.g. from Belle-II) and new lattice calculations. The $\tau\to\pi\pi\nu_\tau$ channel represents a particularly interesting example through the direct comparison of its spectrum and $e^+e^-\to\pi^+\pi^-$ data. Last, the extension of our analysis to strange decays of the tau lepton represents another interesting research line for the future. 


\begin{acknowledgments}
 We thank Mattia Bruno, Bogdan Malaescu, Jorge Martin Camalich, Toni Pich, Jorge Portoles and Pablo Roig for useful discussions. 
Work supported by the European Commission's Horizon 2020 Programme under the Marie Sk\l{}odowska-Curie grant agreements No 690575 and No 674896, 
and the Marie Sk\l{}odowska-Curie Individual Fellowship No 745954 (Tau-SYNERGIES),the Agencia Estatal de Investigaci\'on (AEI, ES), and the European Regional Development Fund (ERDF, EU) [Grants No. FPU14/02990, FPA2014-53631-C2-1-P, FPA2017-84445-P and SEV-2014-0398]. 
This work was supported by the Swedish Research Council grants contract numbers 2015-04089 and 2016-05996 and by the European Research Council (ERC) under the European Union’s Horizon 2020 research and innovation programme (grant agreement No 668679). 
\end{acknowledgments}

\appendix

\section{Inclusive decays calculation}
\label{app:inclusive}

The SM contributions $X_{VV/AA}$ to the inclusive observable in Eq. (8) can be written in the standard form of a contour integral in the complex plane:
\begin{equation}
X_{VV/AA} = \frac{i}{2\pi}\oint_{|s|=s_{0}}\frac{ds}{s_{0}}\omega\left(\frac{s}{s_{0}}\right) \Pi_{VV/AA}^{(1+0)}(s)~,
\label{eq:contour}
\tag{S.1}
\end{equation}
which can be calculated using the Operator Product Expansion (OPE) of the corresponding 2-point correlation function $\Pi_{VV/AA}^{(1+0)}$~\cite{Shifman:1978bx}, that is
\begin{equation}
\Pi_{VV/AA}^{(1+0)}(z) = f^{(pert.)}(\alpha_s;z) + \sum \cO_{2d}/(-z)^d~,
\label{eq:ope}
\tag{S.2}
\end{equation}
The first term is the dominant perturbative contribution, which is equal for vector and axial correlators. We calculate it using $\alpha_{s}(M_{z})=0.1182(12)$ from the lattice \cite{Aoki:2016frl} and the latest calculation of the correlator including $\cO(\alpha_s^4)$ contributions~\cite{Baikov:2008jh}. The contour integration is carried out using both fixed-order~\cite{Braaten:1991qm} and contour-improved~\cite{Pivovarov:1991rh,LeDiberder:1992te} perturbation theory. The difference between them give a subleading error to our constraints.

The second term in~\eref{ope} are small nonperturbative power corrections. In the $V+A$ channel we take them into account in a conservative way through $\cO^{V\!+\!A}_{2d}=0 \pm \Lambda^{2d}\,(d -1)!$. We use $\Lambda\approx 0.4\, \mathrm{GeV}$ as the naive scale of the power corrections, based on the few known vacuum condensates, and the factorial factor accounts for the possible asymptotic behaviour. 
In the $V-A$ channel only the dimension-six contribution is needed for the weight functions we chose. Using the naive estimate above for this contribution would give a very significant error to Eq. (12). Fortunately it is possible to obtain a more precise estimate for this particular term. Combining dispersive methods~\cite{Donoghue:1999ku,Cirigliano:2001qw,Cirigliano:2002jy} with the lattice results on $K\rightarrow \pi\pi$ matrix elements~\cite{Blum:2012uk} we obtain $\cO_{6}^{V\!-\!A}=-0.0042(13)\, \mathrm{GeV}^{6}$.

Using also an OPE for the vector-tensor correlator~\cite{Balitsky:1985aq,Cata:2008zc} we find
\begin{equation}
X_{VT} 
= 
- 48\delta_{n,0} \frac{\langle \bar{q}q \rangle}{s_0\,m_\tau} 
\approx 
0.172(86) \frac{m^2_\tau}{s_0}\delta_{n,0}~.
\tag{S.3}
\end{equation}
for $\omega(x)=x^n$, and using the latest $N_f=2+1$ lattice average for the quark condensate~\cite{FLAG2017}, from Refs.~\cite{Cossu:2016eqs,Durr:2013goa,Bazavov:2010yq,Borsanyi:2012zv,Boyle:2015exm}. We assign a conservative 50\% uncertainty to our simple estimate, and we work with the lower value (see Ref.~\cite{companion} for a more thorough study of this uncertainty).

Using the OPE result in the entire contour in~\eref{contour} introduces a systematic error that is known as Quark-Hadron Duality Violations~\cite{Braaten:1991qm,Chibisov:1996wf,Shifman:2000jv,Cata:2005zj,GonzalezAlonso:2010xf,Boito:2014sta,Pich:2016bdg}. They are typically small for $s_{0}\sim m_{\tau}^{2}$ and for the appropriate weight functions, but its precise  calculation is a nontrivial task. We estimate their size in this work using a conservative $s_{0}-$stability criteria.

\section{SMEFT Lagrangian}
\label{app:SMEFT}
The vertex corrections in Eq. (13) parametrize $W$ boson interactions with quarks and leptons in the SMEFT Lagrangian after electroweak symmetry breaking: 
\begin{equation}
\cL \supset {g_L \over \sqrt 2} W_\mu^+  \left[ 
\delta g_R^{W q_1}  \bar u  \gamma_\mu P_R d
+  \left(1 + \delta g_L^{W\ell}   \right ) \bar \ell  \gamma_\mu P_L \nu_\ell 
\right] + \hc 
\tag{S.4}
\end{equation} 
where $\ell = \tau,e$. 
In the Higgs basis the $\delta g$'s are treated as independent parameters spanning the space of dimension-6 operators; 
in the Warsaw basis they can be expressed as linear combinations of several Wilson coefficients~\cite{deFlorian:2016spz,Falkowski:2017pss}. 
The vertex correction $\delta g_L^{W q_1}$ (parametrizing the $W$ coupling to left-handed up and down quarks)  does not appear in Eq. (13) because its effect cancels in the $\eL^\tau-\eL^e$ difference. 
The four-fermion coefficients $c_i$ are defined by  
\begin{align}
\cL &\supset  \left( [c_{l q}^{(3)}]_{\tau \tau 11} (\bar l \gamma_\mu \sigma^i P_L l)(\bar q \gamma_\mu \sigma^i P_L q) \right.
\nnl
&+ [c_{lequ}]_{\tau \tau 11}  (\bar l P_R e) (\bar q P_R  u) 
\nnl 
&+
 [c_{l e d q}]_{\tau \tau 11} (\bar l P_R  e)(\bar d P_L q)  
\nnl
&+ [c_{lequ}^{(3)}]_{\tau \tau 11}  (\bar l \sigma_{\mu\nu} P_R e) (\bar q \sigma_{\mu\nu} P_R  u)
\left. \right)\frac{1}{v^2}~, 
\tag{S.5}
\end{align}
where $l = (\nu_\tau, \tau)^T$, $q = (u,d)^T$, and $v\approx 246$ GeV is the vacuum expectation value of the Higgs field. 
$[c_{l q}^{(3)}]_{ee11}$ is defined analogously with $l = (\nu_e, e)^T$.
Both in the Higgs and the Warsaw basis the $c_i$ coefficients are independent parameters.

\bibliographystyle{apsrev4-1}
\bibliography{main}

\begin{thebibliography}{81}%
\makeatletter
\providecommand \@ifxundefined [1]{%
 \@ifx{#1\undefined}
}%
\providecommand \@ifnum [1]{%
 \ifnum #1\expandafter \@firstoftwo
 \else \expandafter \@secondoftwo
 \fi
}%
\providecommand \@ifx [1]{%
 \ifx #1\expandafter \@firstoftwo
 \else \expandafter \@secondoftwo
 \fi
}%
\providecommand \natexlab [1]{#1}%
\providecommand \enquote  [1]{``#1''}%
\providecommand \bibnamefont  [1]{#1}%
\providecommand \bibfnamefont [1]{#1}%
\providecommand \citenamefont [1]{#1}%
\providecommand \href@noop [0]{\@secondoftwo}%
\providecommand \href [0]{\begingroup \@sanitize@url \@href}%
\providecommand \@href[1]{\@@startlink{#1}\@@href}%
\providecommand \@@href[1]{\endgroup#1\@@endlink}%
\providecommand \@sanitize@url [0]{\catcode `\\12\catcode `\$12\catcode
  `\&12\catcode `\#12\catcode `\^12\catcode `\_12\catcode `\%12\relax}%
\providecommand \@@startlink[1]{}%
\providecommand \@@endlink[0]{}%
\providecommand \url  [0]{\begingroup\@sanitize@url \@url }%
\providecommand \@url [1]{\endgroup\@href {#1}{\urlprefix }}%
\providecommand \urlprefix  [0]{URL }%
\providecommand \Eprint [0]{\href }%
\providecommand \doibase [0]{http://dx.doi.org/}%
\providecommand \selectlanguage [0]{\@gobble}%
\providecommand \bibinfo  [0]{\@secondoftwo}%
\providecommand \bibfield  [0]{\@secondoftwo}%
\providecommand \translation [1]{[#1]}%
\providecommand \BibitemOpen [0]{}%
\providecommand \bibitemStop [0]{}%
\providecommand \bibitemNoStop [0]{.\EOS\space}%
\providecommand \EOS [0]{\spacefactor3000\relax}%
\providecommand \BibitemShut  [1]{\csname bibitem#1\endcsname}%
\let\auto@bib@innerbib\@empty
\bibitem [{\citenamefont {Pich}(2014)}]{Pich:2013lsa}%
  \BibitemOpen
  \bibfield  {author} {\bibinfo {author} {\bibfnamefont {A.}~\bibnamefont
  {Pich}},\ }\href {\doibase 10.1016/j.ppnp.2013.11.002} {\bibfield  {journal}
  {\bibinfo  {journal} {Prog. Part. Nucl. Phys.}\ }\textbf {\bibinfo {volume}
  {75}},\ \bibinfo {pages} {41} (\bibinfo {year} {2014})},\ \Eprint
  {http://arxiv.org/abs/1310.7922} {arXiv:1310.7922 [hep-ph]} \BibitemShut
  {NoStop}%
\bibitem [{\citenamefont {Schael}\ \emph {et~al.}(2005)\citenamefont {Schael}
  \emph {et~al.}}]{Schael:2005am}%
  \BibitemOpen
  \bibfield  {author} {\bibinfo {author} {\bibfnamefont {S.}~\bibnamefont
  {Schael}} \emph {et~al.} (\bibinfo {collaboration} {ALEPH}),\ }\href
  {\doibase 10.1016/j.physrep.2005.06.007} {\bibfield  {journal} {\bibinfo
  {journal} {Phys. Rept.}\ }\textbf {\bibinfo {volume} {421}},\ \bibinfo
  {pages} {191} (\bibinfo {year} {2005})},\ \Eprint
  {http://arxiv.org/abs/hep-ex/0506072} {arXiv:hep-ex/0506072} \BibitemShut
  {NoStop}%
\bibitem [{\citenamefont {Braaten}\ \emph {et~al.}(1992)\citenamefont
  {Braaten}, \citenamefont {Narison},\ and\ \citenamefont
  {Pich}}]{Braaten:1991qm}%
  \BibitemOpen
  \bibfield  {author} {\bibinfo {author} {\bibfnamefont {E.}~\bibnamefont
  {Braaten}}, \bibinfo {author} {\bibfnamefont {S.}~\bibnamefont {Narison}}, \
  and\ \bibinfo {author} {\bibfnamefont {A.}~\bibnamefont {Pich}},\ }\href
  {\doibase 10.1016/0550-3213(92)90267-F} {\bibfield  {journal} {\bibinfo
  {journal} {Nucl. Phys.}\ }\textbf {\bibinfo {volume} {B373}},\ \bibinfo
  {pages} {581} (\bibinfo {year} {1992})}\BibitemShut {NoStop}%
\bibitem [{\citenamefont {Boito}\ \emph
  {et~al.}(2015{\natexlab{a}})\citenamefont {Boito}, \citenamefont {Golterman},
  \citenamefont {Maltman}, \citenamefont {Osborne},\ and\ \citenamefont
  {Peris}}]{Boito:2014sta}%
  \BibitemOpen
  \bibfield  {author} {\bibinfo {author} {\bibfnamefont {D.}~\bibnamefont
  {Boito}}, \bibinfo {author} {\bibfnamefont {M.}~\bibnamefont {Golterman}},
  \bibinfo {author} {\bibfnamefont {K.}~\bibnamefont {Maltman}}, \bibinfo
  {author} {\bibfnamefont {J.}~\bibnamefont {Osborne}}, \ and\ \bibinfo
  {author} {\bibfnamefont {S.}~\bibnamefont {Peris}},\ }\href {\doibase
  10.1103/PhysRevD.91.034003} {\bibfield  {journal} {\bibinfo  {journal} {Phys.
  Rev.}\ }\textbf {\bibinfo {volume} {D91}},\ \bibinfo {pages} {034003}
  (\bibinfo {year} {2015}{\natexlab{a}})},\ \Eprint
  {http://arxiv.org/abs/1410.3528} {arXiv:1410.3528 [hep-ph]} \BibitemShut
  {NoStop}%
\bibitem [{\citenamefont {Pich}\ and\ \citenamefont
  {Rodriguez-Sanchez}(2016)}]{Pich:2016bdg}%
  \BibitemOpen
  \bibfield  {author} {\bibinfo {author} {\bibfnamefont {A.}~\bibnamefont
  {Pich}}\ and\ \bibinfo {author} {\bibfnamefont {A.}~\bibnamefont
  {Rodriguez-Sanchez}},\ }\href {\doibase 10.1103/PhysRevD.94.034027}
  {\bibfield  {journal} {\bibinfo  {journal} {Phys. Rev.}\ }\textbf {\bibinfo
  {volume} {D94}},\ \bibinfo {pages} {034027} (\bibinfo {year} {2016})},\
  \Eprint {http://arxiv.org/abs/1605.06830} {arXiv:1605.06830 [hep-ph]}
  \BibitemShut {NoStop}%
\bibitem [{\citenamefont {G\'amiz}\ \emph {et~al.}(2003)\citenamefont {G\'amiz}
  \emph {et~al.}}]{Gamiz:2002nu}%
  \BibitemOpen
  \bibfield  {author} {\bibinfo {author} {\bibfnamefont {E.}~\bibnamefont
  {G\'amiz}} \emph {et~al.},\ }\href@noop {} {\bibfield  {journal} {\bibinfo
  {journal} {JHEP}\ }\textbf {\bibinfo {volume} {01}},\ \bibinfo {pages} {060}
  (\bibinfo {year} {2003})},\ \Eprint {http://arxiv.org/abs/hep-ph/0212230}
  {arXiv:hep-ph/0212230} \BibitemShut {NoStop}%
\bibitem [{\citenamefont {G\'amiz}\ \emph {et~al.}(2005)\citenamefont {G\'amiz}
  \emph {et~al.}}]{Gamiz:2004ar}%
  \BibitemOpen
  \bibfield  {author} {\bibinfo {author} {\bibfnamefont {E.}~\bibnamefont
  {G\'amiz}} \emph {et~al.},\ }\href {\doibase 10.1103/PhysRevLett.94.011803}
  {\bibfield  {journal} {\bibinfo  {journal} {Phys. Rev. Lett.}\ }\textbf
  {\bibinfo {volume} {94}},\ \bibinfo {pages} {011803} (\bibinfo {year}
  {2005})},\ \Eprint {http://arxiv.org/abs/hep-ph/0408044}
  {arXiv:hep-ph/0408044} \BibitemShut {NoStop}%
\bibitem [{\citenamefont {Boito}\ \emph
  {et~al.}(2015{\natexlab{b}})\citenamefont {Boito}, \citenamefont {Francis},
  \citenamefont {Golterman}, \citenamefont {Hudspith}, \citenamefont {Lewis},
  \citenamefont {Maltman},\ and\ \citenamefont {Peris}}]{Boito:2015fra}%
  \BibitemOpen
  \bibfield  {author} {\bibinfo {author} {\bibfnamefont {D.}~\bibnamefont
  {Boito}}, \bibinfo {author} {\bibfnamefont {A.}~\bibnamefont {Francis}},
  \bibinfo {author} {\bibfnamefont {M.}~\bibnamefont {Golterman}}, \bibinfo
  {author} {\bibfnamefont {R.}~\bibnamefont {Hudspith}}, \bibinfo {author}
  {\bibfnamefont {R.}~\bibnamefont {Lewis}}, \bibinfo {author} {\bibfnamefont
  {K.}~\bibnamefont {Maltman}}, \ and\ \bibinfo {author} {\bibfnamefont
  {S.}~\bibnamefont {Peris}},\ }\href {\doibase 10.1103/PhysRevD.92.114501}
  {\bibfield  {journal} {\bibinfo  {journal} {Phys. Rev.}\ }\textbf {\bibinfo
  {volume} {D92}},\ \bibinfo {pages} {114501} (\bibinfo {year}
  {2015}{\natexlab{b}})},\ \Eprint {http://arxiv.org/abs/1503.03450}
  {arXiv:1503.03450 [hep-ph]} \BibitemShut {NoStop}%
\bibitem [{\citenamefont {Gonz\'alez-Alonso}\ \emph {et~al.}(2016)\citenamefont
  {Gonz\'alez-Alonso}, \citenamefont {Pich},\ and\ \citenamefont
  {Rodr\'iguez-S\'anchez}}]{Rodriguez-Sanchez:2016jvw}%
  \BibitemOpen
  \bibfield  {author} {\bibinfo {author} {\bibfnamefont {M.}~\bibnamefont
  {Gonz\'alez-Alonso}}, \bibinfo {author} {\bibfnamefont {A.}~\bibnamefont
  {Pich}}, \ and\ \bibinfo {author} {\bibfnamefont {A.}~\bibnamefont
  {Rodr\'iguez-S\'anchez}},\ }\href {\doibase 10.1103/PhysRevD.94.014017}
  {\bibfield  {journal} {\bibinfo  {journal} {Phys. Rev.}\ }\textbf {\bibinfo
  {volume} {D94}},\ \bibinfo {pages} {014017} (\bibinfo {year} {2016})},\
  \Eprint {http://arxiv.org/abs/1602.06112} {arXiv:1602.06112 [hep-ph]}
  \BibitemShut {NoStop}%
\bibitem [{\citenamefont {Aoki}\ \emph {et~al.}(2017)\citenamefont {Aoki} \emph
  {et~al.}}]{Aoki:2016frl}%
  \BibitemOpen
  \bibfield  {author} {\bibinfo {author} {\bibfnamefont {S.}~\bibnamefont
  {Aoki}} \emph {et~al.},\ }\href {\doibase 10.1140/epjc/s10052-016-4509-7}
  {\bibfield  {journal} {\bibinfo  {journal} {Eur. Phys. J.}\ }\textbf
  {\bibinfo {volume} {C77}},\ \bibinfo {pages} {112} (\bibinfo {year}
  {2017})},\ \Eprint {http://arxiv.org/abs/1607.00299} {arXiv:1607.00299
  [hep-lat]} \BibitemShut {NoStop}%
\bibitem [{\citenamefont {Bernard}\ \emph {et~al.}(2008)\citenamefont
  {Bernard}, \citenamefont {Oertel}, \citenamefont {Passemar},\ and\
  \citenamefont {Stern}}]{Bernard:2007cf}%
  \BibitemOpen
  \bibfield  {author} {\bibinfo {author} {\bibfnamefont {V.}~\bibnamefont
  {Bernard}}, \bibinfo {author} {\bibfnamefont {M.}~\bibnamefont {Oertel}},
  \bibinfo {author} {\bibfnamefont {E.}~\bibnamefont {Passemar}}, \ and\
  \bibinfo {author} {\bibfnamefont {J.}~\bibnamefont {Stern}},\ }\href
  {\doibase 10.1088/1126-6708/2008/01/015} {\bibfield  {journal} {\bibinfo
  {journal} {JHEP}\ }\textbf {\bibinfo {volume} {01}},\ \bibinfo {pages} {015}
  (\bibinfo {year} {2008})},\ \Eprint {http://arxiv.org/abs/0707.4194}
  {arXiv:0707.4194 [hep-ph]} \BibitemShut {NoStop}%
\bibitem [{\citenamefont {Garc\'es}\ \emph {et~al.}(2017)\citenamefont
  {Garc\'es}, \citenamefont {Hern\'andez~Villanueva}, \citenamefont
  {L\'opez~Castro},\ and\ \citenamefont {Roig}}]{Garces:2017jpz}%
  \BibitemOpen
  \bibfield  {author} {\bibinfo {author} {\bibfnamefont {E.~A.}\ \bibnamefont
  {Garc\'es}}, \bibinfo {author} {\bibfnamefont {M.}~\bibnamefont
  {Hern\'andez~Villanueva}}, \bibinfo {author} {\bibfnamefont {G.}~\bibnamefont
  {L\'opez~Castro}}, \ and\ \bibinfo {author} {\bibfnamefont {P.}~\bibnamefont
  {Roig}},\ }\href {\doibase 10.1007/JHEP12(2017)027} {\bibfield  {journal}
  {\bibinfo  {journal} {JHEP}\ }\textbf {\bibinfo {volume} {12}},\ \bibinfo
  {pages} {027} (\bibinfo {year} {2017})},\ \Eprint
  {http://arxiv.org/abs/1708.07802} {arXiv:1708.07802 [hep-ph]} \BibitemShut
  {NoStop}%
\bibitem [{\citenamefont {Cirigliano}\ \emph {et~al.}(2010)\citenamefont
  {Cirigliano}, \citenamefont {Jenkins},\ and\ \citenamefont
  {Gonzalez-Alonso}}]{Cirigliano:2009wk}%
  \BibitemOpen
  \bibfield  {author} {\bibinfo {author} {\bibfnamefont {V.}~\bibnamefont
  {Cirigliano}}, \bibinfo {author} {\bibfnamefont {J.}~\bibnamefont {Jenkins}},
  \ and\ \bibinfo {author} {\bibfnamefont {M.}~\bibnamefont
  {Gonzalez-Alonso}},\ }\href {\doibase 10.1016/j.nuclphysb.2009.12.020}
  {\bibfield  {journal} {\bibinfo  {journal} {Nucl. Phys.}\ }\textbf {\bibinfo
  {volume} {B830}},\ \bibinfo {pages} {95} (\bibinfo {year} {2010})},\ \Eprint
  {http://arxiv.org/abs/0908.1754} {arXiv:0908.1754 [hep-ph]} \BibitemShut
  {NoStop}%
\bibitem [{\citenamefont {Cirigliano}\ \emph {et~al.}(2013)\citenamefont
  {Cirigliano}, \citenamefont {Gonzalez-Alonso},\ and\ \citenamefont
  {Graesser}}]{Cirigliano:2012ab}%
  \BibitemOpen
  \bibfield  {author} {\bibinfo {author} {\bibfnamefont {V.}~\bibnamefont
  {Cirigliano}}, \bibinfo {author} {\bibfnamefont {M.}~\bibnamefont
  {Gonzalez-Alonso}}, \ and\ \bibinfo {author} {\bibfnamefont {M.~L.}\
  \bibnamefont {Graesser}},\ }\href {\doibase 10.1007/JHEP02(2013)046}
  {\bibfield  {journal} {\bibinfo  {journal} {JHEP}\ }\textbf {\bibinfo
  {volume} {02}},\ \bibinfo {pages} {046} (\bibinfo {year} {2013})},\ \Eprint
  {http://arxiv.org/abs/1210.4553} {arXiv:1210.4553 [hep-ph]} \BibitemShut
  {NoStop}%
\bibitem [{\citenamefont {Gonzalez-Alonso}\ \emph {et~al.}(2019)\citenamefont
  {Gonzalez-Alonso}, \citenamefont {Naviliat-Cuncic},\ and\ \citenamefont
  {Severijns}}]{Gonzalez-Alonso:2018omy}%
  \BibitemOpen
  \bibfield  {author} {\bibinfo {author} {\bibfnamefont {M.}~\bibnamefont
  {Gonzalez-Alonso}}, \bibinfo {author} {\bibfnamefont {O.}~\bibnamefont
  {Naviliat-Cuncic}}, \ and\ \bibinfo {author} {\bibfnamefont {N.}~\bibnamefont
  {Severijns}},\ }\href@noop {} {\bibfield  {journal} {\bibinfo  {journal}
  {Prog. Part. Nucl. Phys.}\ }\textbf {\bibinfo {volume} {104}},\ \bibinfo
  {pages} {165} (\bibinfo {year} {2019})},\ \Eprint
  {http://arxiv.org/abs/1803.08732} {arXiv:1803.08732 [hep-ph]} \BibitemShut
  {NoStop}%
\bibitem [{\citenamefont {Patrignani}\ \emph {et~al.}(2016)\citenamefont
  {Patrignani} \emph {et~al.}}]{Patrignani:2016xqp}%
  \BibitemOpen
  \bibfield  {author} {\bibinfo {author} {\bibfnamefont {C.}~\bibnamefont
  {Patrignani}} \emph {et~al.} (\bibinfo {collaboration} {Particle Data
  Group}),\ }\href {\doibase 10.1088/1674-1137/40/10/100001} {\bibfield
  {journal} {\bibinfo  {journal} {Chin. Phys.}\ }\textbf {\bibinfo {volume}
  {C40}},\ \bibinfo {pages} {100001} (\bibinfo {year} {2016})}\BibitemShut
  {NoStop}%
\bibitem [{\citenamefont {Decker}\ and\ \citenamefont
  {Finkemeier}(1995)}]{Decker:1994ea}%
  \BibitemOpen
  \bibfield  {author} {\bibinfo {author} {\bibfnamefont {R.}~\bibnamefont
  {Decker}}\ and\ \bibinfo {author} {\bibfnamefont {M.}~\bibnamefont
  {Finkemeier}},\ }\href {\doibase 10.1016/0550-3213(95)00597-L} {\bibfield
  {journal} {\bibinfo  {journal} {Nucl. Phys.}\ }\textbf {\bibinfo {volume}
  {B438}},\ \bibinfo {pages} {17} (\bibinfo {year} {1995})},\ \Eprint
  {http://arxiv.org/abs/hep-ph/9403385} {arXiv:hep-ph/9403385 [hep-ph]}
  \BibitemShut {NoStop}%
\bibitem [{\citenamefont {Cirigliano}\ and\ \citenamefont
  {Rosell}(2007)}]{Cirigliano:2007xi}%
  \BibitemOpen
  \bibfield  {author} {\bibinfo {author} {\bibfnamefont {V.}~\bibnamefont
  {Cirigliano}}\ and\ \bibinfo {author} {\bibfnamefont {I.}~\bibnamefont
  {Rosell}},\ }\href {\doibase 10.1103/PhysRevLett.99.231801} {\bibfield
  {journal} {\bibinfo  {journal} {Phys.Rev.Lett.}\ }\textbf {\bibinfo {volume}
  {99}},\ \bibinfo {pages} {231801} (\bibinfo {year} {2007})},\ \Eprint
  {http://arxiv.org/abs/0707.3439} {arXiv:0707.3439 [hep-ph]} \BibitemShut
  {NoStop}%
\bibitem [{\citenamefont {Rosner}\ \emph {et~al.}(2015)\citenamefont {Rosner},
  \citenamefont {Stone},\ and\ \citenamefont {Van~de Water}}]{Rosner:2015wva}%
  \BibitemOpen
  \bibfield  {author} {\bibinfo {author} {\bibfnamefont {J.~L.}\ \bibnamefont
  {Rosner}}, \bibinfo {author} {\bibfnamefont {S.}~\bibnamefont {Stone}}, \
  and\ \bibinfo {author} {\bibfnamefont {R.~S.}\ \bibnamefont {Van~de Water}},\
  }\href@noop {} {\bibfield  {journal} {\bibinfo  {journal} {Submitted to:
  Particle Data Book}\ } (\bibinfo {year} {2015})},\ \Eprint
  {http://arxiv.org/abs/1509.02220} {arXiv:1509.02220 [hep-ph]} \BibitemShut
  {NoStop}%
\bibitem [{\citenamefont {Aoki}\ \emph {et~al.}()\citenamefont {Aoki} \emph
  {et~al.}}]{FLAG2017}%
  \BibitemOpen
  \bibfield  {author} {\bibinfo {author} {\bibfnamefont {S.}~\bibnamefont
  {Aoki}} \emph {et~al.},\ }\href@noop {} {\ }\bibinfo {note}
  {Http://flag.unibe.ch/ (Dec. 2017 version)}\BibitemShut {NoStop}%
\bibitem [{\citenamefont {Blum}\ \emph {et~al.}(2016)\citenamefont {Blum} \emph
  {et~al.}}]{Blum:2014tka}%
  \BibitemOpen
  \bibfield  {author} {\bibinfo {author} {\bibfnamefont {T.}~\bibnamefont
  {Blum}} \emph {et~al.} (\bibinfo {collaboration} {RBC, UKQCD}),\ }\href
  {\doibase 10.1103/PhysRevD.93.074505} {\bibfield  {journal} {\bibinfo
  {journal} {Phys. Rev.}\ }\textbf {\bibinfo {volume} {D93}},\ \bibinfo {pages}
  {074505} (\bibinfo {year} {2016})},\ \Eprint {http://arxiv.org/abs/1411.7017}
  {arXiv:1411.7017 [hep-lat]} \BibitemShut {NoStop}%
\bibitem [{\citenamefont {Bazavov}\ \emph
  {et~al.}(2010{\natexlab{a}})\citenamefont {Bazavov} \emph
  {et~al.}}]{Bazavov:2010hj}%
  \BibitemOpen
  \bibfield  {author} {\bibinfo {author} {\bibfnamefont {A.}~\bibnamefont
  {Bazavov}} \emph {et~al.} (\bibinfo {collaboration} {MILC}),\ }\bibfield
  {booktitle} {\emph {\bibinfo {booktitle} {{Proceedings, 28th International
  Symposium on Lattice field theory (Lattice 2010): Villasimius, Italy, June
  14-19, 2010}}},\ }\href {\doibase 10.22323/1.105.0074} {\bibfield  {journal}
  {\bibinfo  {journal} {PoS}\ }\textbf {\bibinfo {volume} {LATTICE2010}},\
  \bibinfo {pages} {074} (\bibinfo {year} {2010}{\natexlab{a}})},\ \Eprint
  {http://arxiv.org/abs/1012.0868} {arXiv:1012.0868 [hep-lat]} \BibitemShut
  {NoStop}%
\bibitem [{\citenamefont {Follana}\ \emph {et~al.}(2008)\citenamefont
  {Follana}, \citenamefont {Davies}, \citenamefont {Lepage},\ and\
  \citenamefont {Shigemitsu}}]{Follana:2007uv}%
  \BibitemOpen
  \bibfield  {author} {\bibinfo {author} {\bibfnamefont {E.}~\bibnamefont
  {Follana}}, \bibinfo {author} {\bibfnamefont {C.~T.~H.}\ \bibnamefont
  {Davies}}, \bibinfo {author} {\bibfnamefont {G.~P.}\ \bibnamefont {Lepage}},
  \ and\ \bibinfo {author} {\bibfnamefont {J.}~\bibnamefont {Shigemitsu}}
  (\bibinfo {collaboration} {HPQCD, UKQCD}),\ }\href {\doibase
  10.1103/PhysRevLett.100.062002} {\bibfield  {journal} {\bibinfo  {journal}
  {Phys. Rev. Lett.}\ }\textbf {\bibinfo {volume} {100}},\ \bibinfo {pages}
  {062002} (\bibinfo {year} {2008})},\ \Eprint {http://arxiv.org/abs/0706.1726}
  {arXiv:0706.1726 [hep-lat]} \BibitemShut {NoStop}%
\bibitem [{\citenamefont {Gonzalez-Alonso}\ and\ \citenamefont
  {Martin~Camalich}(2016)}]{Gonzalez-Alonso:2016etj}%
  \BibitemOpen
  \bibfield  {author} {\bibinfo {author} {\bibfnamefont {M.}~\bibnamefont
  {Gonzalez-Alonso}}\ and\ \bibinfo {author} {\bibfnamefont {J.}~\bibnamefont
  {Martin~Camalich}},\ }\href {\doibase 10.1007/JHEP12(2016)052} {\bibfield
  {journal} {\bibinfo  {journal} {JHEP}\ }\textbf {\bibinfo {volume} {12}},\
  \bibinfo {pages} {052} (\bibinfo {year} {2016})},\ \Eprint
  {http://arxiv.org/abs/1605.07114} {arXiv:1605.07114 [hep-ph]} \BibitemShut
  {NoStop}%
\bibitem [{\citenamefont {Davier}\ \emph {et~al.}(2013)\citenamefont {Davier},
  \citenamefont {Hoecker}, \citenamefont {Malaescu}, \citenamefont {Yuan},\
  and\ \citenamefont {Zhang}}]{Davier:2013sfa}%
  \BibitemOpen
  \bibfield  {author} {\bibinfo {author} {\bibfnamefont {M.}~\bibnamefont
  {Davier}}, \bibinfo {author} {\bibfnamefont {A.}~\bibnamefont {Hoecker}},
  \bibinfo {author} {\bibfnamefont {B.}~\bibnamefont {Malaescu}}, \bibinfo
  {author} {\bibfnamefont {C.}~\bibnamefont {Yuan}}, \ and\ \bibinfo {author}
  {\bibfnamefont {Z.}~\bibnamefont {Zhang}},\ }\href@noop {} {\  (\bibinfo
  {year} {2013})},\ \Eprint {http://arxiv.org/abs/1312.1501} {arXiv:1312.1501
  [hep-ex]} \BibitemShut {NoStop}%
\bibitem [{\citenamefont {Davier}\ \emph {et~al.}(2017)\citenamefont {Davier},
  \citenamefont {Hoecker}, \citenamefont {Malaescu},\ and\ \citenamefont
  {Zhang}}]{Davier:2017zfy}%
  \BibitemOpen
  \bibfield  {author} {\bibinfo {author} {\bibfnamefont {M.}~\bibnamefont
  {Davier}}, \bibinfo {author} {\bibfnamefont {A.}~\bibnamefont {Hoecker}},
  \bibinfo {author} {\bibfnamefont {B.}~\bibnamefont {Malaescu}}, \ and\
  \bibinfo {author} {\bibfnamefont {Z.}~\bibnamefont {Zhang}},\ }\href@noop {}
  {\  (\bibinfo {year} {2017})},\ \Eprint {http://arxiv.org/abs/1706.09436}
  {arXiv:1706.09436 [hep-ph]} \BibitemShut {NoStop}%
\bibitem [{\citenamefont {Keshavarzi}\ \emph {et~al.}(2018)\citenamefont
  {Keshavarzi}, \citenamefont {Nomura},\ and\ \citenamefont
  {Teubner}}]{Keshavarzi:2018mgv}%
  \BibitemOpen
  \bibfield  {author} {\bibinfo {author} {\bibfnamefont {A.}~\bibnamefont
  {Keshavarzi}}, \bibinfo {author} {\bibfnamefont {D.}~\bibnamefont {Nomura}},
  \ and\ \bibinfo {author} {\bibfnamefont {T.}~\bibnamefont {Teubner}},\
  }\href@noop {} {\  (\bibinfo {year} {2018})},\ \Eprint
  {http://arxiv.org/abs/1802.02995} {arXiv:1802.02995 [hep-ph]} \BibitemShut
  {NoStop}%
\bibitem [{\citenamefont {Cirigliano}\ \emph {et~al.}(2018)\citenamefont
  {Cirigliano}, \citenamefont {Crivellin},\ and\ \citenamefont
  {Hoferichter}}]{Cirigliano:2017tqn}%
  \BibitemOpen
  \bibfield  {author} {\bibinfo {author} {\bibfnamefont {V.}~\bibnamefont
  {Cirigliano}}, \bibinfo {author} {\bibfnamefont {A.}~\bibnamefont
  {Crivellin}}, \ and\ \bibinfo {author} {\bibfnamefont {M.}~\bibnamefont
  {Hoferichter}},\ }\href {\doibase 10.1103/PhysRevLett.120.141803} {\bibfield
  {journal} {\bibinfo  {journal} {Phys. Rev. Lett.}\ }\textbf {\bibinfo
  {volume} {120}},\ \bibinfo {pages} {141803} (\bibinfo {year} {2018})},\
  \Eprint {http://arxiv.org/abs/1712.06595} {arXiv:1712.06595 [hep-ph]}
  \BibitemShut {NoStop}%
\bibitem [{\citenamefont {Miranda}\ and\ \citenamefont
  {Roig}(2018)}]{Miranda:2018cpf}%
  \BibitemOpen
  \bibfield  {author} {\bibinfo {author} {\bibfnamefont {J.~A.}\ \bibnamefont
  {Miranda}}\ and\ \bibinfo {author} {\bibfnamefont {P.}~\bibnamefont {Roig}},\
  }\href@noop {} {\  (\bibinfo {year} {2018})},\ \Eprint
  {http://arxiv.org/abs/1806.09547} {arXiv:1806.09547 [hep-ph]} \BibitemShut
  {NoStop}%
\bibitem [{\citenamefont {Ecker}\ \emph {et~al.}(1989)\citenamefont {Ecker},
  \citenamefont {Gasser}, \citenamefont {Pich},\ and\ \citenamefont
  {de~Rafael}}]{Ecker:1988te}%
  \BibitemOpen
  \bibfield  {author} {\bibinfo {author} {\bibfnamefont {G.}~\bibnamefont
  {Ecker}}, \bibinfo {author} {\bibfnamefont {J.}~\bibnamefont {Gasser}},
  \bibinfo {author} {\bibfnamefont {A.}~\bibnamefont {Pich}}, \ and\ \bibinfo
  {author} {\bibfnamefont {E.}~\bibnamefont {de~Rafael}},\ }\href {\doibase
  10.1016/0550-3213(89)90346-5} {\bibfield  {journal} {\bibinfo  {journal}
  {Nucl. Phys.}\ }\textbf {\bibinfo {volume} {B321}},\ \bibinfo {pages} {311}
  (\bibinfo {year} {1989})}\BibitemShut {NoStop}%
\bibitem [{\citenamefont {Baum}\ \emph {et~al.}(2011)\citenamefont {Baum},
  \citenamefont {Lubicz}, \citenamefont {Martinelli}, \citenamefont {Orifici},\
  and\ \citenamefont {Simula}}]{Baum:2011rm}%
  \BibitemOpen
  \bibfield  {author} {\bibinfo {author} {\bibfnamefont {I.}~\bibnamefont
  {Baum}}, \bibinfo {author} {\bibfnamefont {V.}~\bibnamefont {Lubicz}},
  \bibinfo {author} {\bibfnamefont {G.}~\bibnamefont {Martinelli}}, \bibinfo
  {author} {\bibfnamefont {L.}~\bibnamefont {Orifici}}, \ and\ \bibinfo
  {author} {\bibfnamefont {S.}~\bibnamefont {Simula}},\ }\href {\doibase
  10.1103/PhysRevD.84.074503} {\bibfield  {journal} {\bibinfo  {journal} {Phys.
  Rev.}\ }\textbf {\bibinfo {volume} {D84}},\ \bibinfo {pages} {074503}
  (\bibinfo {year} {2011})},\ \Eprint {http://arxiv.org/abs/1108.1021}
  {arXiv:1108.1021 [hep-lat]} \BibitemShut {NoStop}%
\bibitem [{\citenamefont {Mateu}\ and\ \citenamefont
  {Portol\'es}(2007)}]{Mateu:2007tr}%
  \BibitemOpen
  \bibfield  {author} {\bibinfo {author} {\bibfnamefont {V.}~\bibnamefont
  {Mateu}}\ and\ \bibinfo {author} {\bibfnamefont {J.}~\bibnamefont
  {Portol\'es}},\ }\href {\doibase 10.1140/epjc/s10052-007-0393-5} {\bibfield
  {journal} {\bibinfo  {journal} {Eur. Phys. J.}\ }\textbf {\bibinfo {volume}
  {C52}},\ \bibinfo {pages} {325} (\bibinfo {year} {2007})},\ \Eprint
  {http://arxiv.org/abs/0706.1039} {arXiv:0706.1039 [hep-ph]} \BibitemShut
  {NoStop}%
\bibitem [{\citenamefont {Cata}\ and\ \citenamefont
  {Mateu}(2008)}]{Cata:2008zc}%
  \BibitemOpen
  \bibfield  {author} {\bibinfo {author} {\bibfnamefont {O.}~\bibnamefont
  {Cata}}\ and\ \bibinfo {author} {\bibfnamefont {V.}~\bibnamefont {Mateu}},\
  }\href {\doibase 10.1103/PhysRevD.77.116009} {\bibfield  {journal} {\bibinfo
  {journal} {Phys. Rev.}\ }\textbf {\bibinfo {volume} {D77}},\ \bibinfo {pages}
  {116009} (\bibinfo {year} {2008})},\ \Eprint {http://arxiv.org/abs/0801.4374}
  {arXiv:0801.4374 [hep-ph]} \BibitemShut {NoStop}%
\bibitem [{\citenamefont {Cirigliano}\ \emph {et~al.}(2019)\citenamefont
  {Cirigliano}, \citenamefont {Falkowski}, \citenamefont
  {Gonz\'{a}lez-Alonso},\ and\ \citenamefont
  {Rodr\'iguez-S\'anchez}}]{companion}%
  \BibitemOpen
  \bibfield  {author} {\bibinfo {author} {\bibfnamefont {V.}~\bibnamefont
  {Cirigliano}}, \bibinfo {author} {\bibfnamefont {A.}~\bibnamefont
  {Falkowski}}, \bibinfo {author} {\bibfnamefont {M.}~\bibnamefont
  {Gonz\'{a}lez-Alonso}}, \ and\ \bibinfo {author} {\bibfnamefont
  {A.}~\bibnamefont {Rodr\'iguez-S\'anchez}},\ }\href@noop {} {\enquote
  {\bibinfo {title} {{\it Model-independent analysis of non-standard effects in
  (non-)strange Hadronic Tau Decays }},}\ } (\bibinfo {year} {to appear,
  2019})\BibitemShut {NoStop}%
\bibitem [{\citenamefont {Bruno}\ \emph {et~al.}(2018)\citenamefont {Bruno},
  \citenamefont {Izubuchi}, \citenamefont {Lehner},\ and\ \citenamefont
  {Meyer}}]{Bruno:2018ono}%
  \BibitemOpen
  \bibfield  {author} {\bibinfo {author} {\bibfnamefont {M.}~\bibnamefont
  {Bruno}}, \bibinfo {author} {\bibfnamefont {T.}~\bibnamefont {Izubuchi}},
  \bibinfo {author} {\bibfnamefont {C.}~\bibnamefont {Lehner}}, \ and\ \bibinfo
  {author} {\bibfnamefont {A.}~\bibnamefont {Meyer}},\ }\bibfield  {booktitle}
  {\emph {\bibinfo {booktitle} {{36th International Symposium on Lattice Field
  Theory (Lattice 2018) East Lansing, MI, United States, July 22-28, 2018}}},\
  }\href@noop {} {\bibfield  {journal} {\bibinfo  {journal} {PoS}\ }\textbf
  {\bibinfo {volume} {LATTICE2018}},\ \bibinfo {pages} {135} (\bibinfo {year}
  {2018})},\ \Eprint {http://arxiv.org/abs/1811.00508} {arXiv:1811.00508
  [hep-lat]} \BibitemShut {NoStop}%
\bibitem [{\citenamefont {del Amo~Sanchez}\ \emph {et~al.}(2011)\citenamefont
  {del Amo~Sanchez} \emph {et~al.}}]{delAmoSanchez:2010pc}%
  \BibitemOpen
  \bibfield  {author} {\bibinfo {author} {\bibfnamefont {P.}~\bibnamefont {del
  Amo~Sanchez}} \emph {et~al.} (\bibinfo {collaboration} {BaBar}),\ }\href
  {\doibase 10.1103/PhysRevD.83.032002} {\bibfield  {journal} {\bibinfo
  {journal} {Phys. Rev.}\ }\textbf {\bibinfo {volume} {D83}},\ \bibinfo {pages}
  {032002} (\bibinfo {year} {2011})},\ \Eprint {http://arxiv.org/abs/1011.3917}
  {arXiv:1011.3917 [hep-ex]} \BibitemShut {NoStop}%
\bibitem [{\citenamefont {Escribano}\ \emph {et~al.}(2016)\citenamefont
  {Escribano}, \citenamefont {Gonzalez-Solis},\ and\ \citenamefont
  {Roig}}]{Escribano:2016ntp}%
  \BibitemOpen
  \bibfield  {author} {\bibinfo {author} {\bibfnamefont {R.}~\bibnamefont
  {Escribano}}, \bibinfo {author} {\bibfnamefont {S.}~\bibnamefont
  {Gonzalez-Solis}}, \ and\ \bibinfo {author} {\bibfnamefont {P.}~\bibnamefont
  {Roig}},\ }\href {\doibase 10.1103/PhysRevD.94.034008} {\bibfield  {journal}
  {\bibinfo  {journal} {Phys. Rev.}\ }\textbf {\bibinfo {volume} {D94}},\
  \bibinfo {pages} {034008} (\bibinfo {year} {2016})},\ \Eprint
  {http://arxiv.org/abs/1601.03989} {arXiv:1601.03989 [hep-ph]} \BibitemShut
  {NoStop}%
\bibitem [{\citenamefont {Roig}()}]{Roig:private}%
  \BibitemOpen
  \bibfield  {author} {\bibinfo {author} {\bibfnamefont {P.}~\bibnamefont
  {Roig}},\ }\href@noop {} {}\bibinfo {howpublished} {Private
  communication}\BibitemShut {NoStop}%
\bibitem [{\citenamefont {Kou}\ \emph {et~al.}(2018)\citenamefont {Kou} \emph
  {et~al.}}]{Kou:2018nap}%
  \BibitemOpen
  \bibfield  {author} {\bibinfo {author} {\bibfnamefont {E.}~\bibnamefont
  {Kou}} \emph {et~al.},\ }\href@noop {} {\  (\bibinfo {year} {2018})},\
  \Eprint {http://arxiv.org/abs/1808.10567} {arXiv:1808.10567 [hep-ex]}
  \BibitemShut {NoStop}%
\bibitem [{\citenamefont {Hayasaka}(2009)}]{Hayasaka:2009zz}%
  \BibitemOpen
  \bibfield  {author} {\bibinfo {author} {\bibfnamefont {K.}~\bibnamefont
  {Hayasaka}} (\bibinfo {collaboration} {Belle}),\ }\bibfield  {booktitle}
  {\emph {\bibinfo {booktitle} {{Proceedings, Europhysics Conference on High
  energy physics (EPS-HEP 2009): Cracow, Poland, July 16-22, 2009}}},\
  }\href@noop {} {\bibfield  {journal} {\bibinfo  {journal} {PoS}\ }\textbf
  {\bibinfo {volume} {EPS-HEP2009}},\ \bibinfo {pages} {374} (\bibinfo {year}
  {2009})}\BibitemShut {NoStop}%
\bibitem [{\citenamefont {Gonzalez-Alonso}(2010)}]{Gonzalez-Alonso:2010vnm}%
  \BibitemOpen
  \bibfield  {author} {\bibinfo {author} {\bibfnamefont {M.}~\bibnamefont
  {Gonzalez-Alonso}},\ }\emph {\bibinfo {title} {{Low-energy tests of the
  Standard Model}}},\ \href
  {https://www.educacion.gob.es/teseo/mostrarRef.do?ref=884247} {Ph.D.
  thesis},\ \bibinfo  {school} {Valencia U.} (\bibinfo {year}
  {2010})\BibitemShut {NoStop}%
\bibitem [{\citenamefont {Rodriguez-Sanchez}(2018)}]{Rodriguez-Sanchez:2018}%
  \BibitemOpen
  \bibfield  {author} {\bibinfo {author} {\bibfnamefont {A.}~\bibnamefont
  {Rodriguez-Sanchez}},\ }\emph {\bibinfo {title} {{Precision Physics in
  Hadronic Tau Decays}}},\ \href@noop {} {Ph.D. thesis},\ \bibinfo  {school}
  {Valencia U.} (\bibinfo {year} {2018})\BibitemShut {NoStop}%
\bibitem [{\citenamefont {Buchmuller}\ and\ \citenamefont
  {Wyler}(1986)}]{Buchmuller:1985jz}%
  \BibitemOpen
  \bibfield  {author} {\bibinfo {author} {\bibfnamefont {W.}~\bibnamefont
  {Buchmuller}}\ and\ \bibinfo {author} {\bibfnamefont {D.}~\bibnamefont
  {Wyler}},\ }\href {\doibase 10.1016/0550-3213(86)90262-2} {\bibfield
  {journal} {\bibinfo  {journal} {Nucl. Phys.}\ }\textbf {\bibinfo {volume}
  {B268}},\ \bibinfo {pages} {621} (\bibinfo {year} {1986})}\BibitemShut
  {NoStop}%
\bibitem [{\citenamefont {Grzadkowski}\ \emph {et~al.}(2010)\citenamefont
  {Grzadkowski}, \citenamefont {Iskrzynski}, \citenamefont {Misiak},\ and\
  \citenamefont {Rosiek}}]{Grzadkowski:2010es}%
  \BibitemOpen
  \bibfield  {author} {\bibinfo {author} {\bibfnamefont {B.}~\bibnamefont
  {Grzadkowski}}, \bibinfo {author} {\bibfnamefont {M.}~\bibnamefont
  {Iskrzynski}}, \bibinfo {author} {\bibfnamefont {M.}~\bibnamefont {Misiak}},
  \ and\ \bibinfo {author} {\bibfnamefont {J.}~\bibnamefont {Rosiek}},\ }\href
  {\doibase 10.1007/JHEP10(2010)085} {\bibfield  {journal} {\bibinfo  {journal}
  {JHEP}\ }\textbf {\bibinfo {volume} {1010}},\ \bibinfo {pages} {085}
  (\bibinfo {year} {2010})},\ \Eprint {http://arxiv.org/abs/1008.4884}
  {arXiv:1008.4884 [hep-ph]} \BibitemShut {NoStop}%
\bibitem [{\citenamefont {de~Florian}\ \emph {et~al.}(2016)\citenamefont
  {de~Florian} \emph {et~al.}}]{deFlorian:2016spz}%
  \BibitemOpen
  \bibfield  {author} {\bibinfo {author} {\bibfnamefont {D.}~\bibnamefont
  {de~Florian}} \emph {et~al.} (\bibinfo {collaboration} {LHC Higgs Cross
  Section Working Group}),\ }\href {\doibase 10.23731/CYRM-2017-002} {\
  (\bibinfo {year} {2016}),\ 10.23731/CYRM-2017-002},\ \Eprint
  {http://arxiv.org/abs/1610.07922} {arXiv:1610.07922 [hep-ph]} \BibitemShut
  {NoStop}%
\bibitem [{\citenamefont {Falkowski}\ \emph {et~al.}(2017)\citenamefont
  {Falkowski}, \citenamefont {Gonzalez-Alonso},\ and\ \citenamefont
  {Mimouni}}]{Falkowski:2017pss}%
  \BibitemOpen
  \bibfield  {author} {\bibinfo {author} {\bibfnamefont {A.}~\bibnamefont
  {Falkowski}}, \bibinfo {author} {\bibfnamefont {M.}~\bibnamefont
  {Gonzalez-Alonso}}, \ and\ \bibinfo {author} {\bibfnamefont {K.}~\bibnamefont
  {Mimouni}},\ }\href {\doibase 10.1007/JHEP08(2017)123} {\bibfield  {journal}
  {\bibinfo  {journal} {JHEP}\ }\textbf {\bibinfo {volume} {08}},\ \bibinfo
  {pages} {123} (\bibinfo {year} {2017})},\ \Eprint
  {http://arxiv.org/abs/1706.03783} {arXiv:1706.03783 [hep-ph]} \BibitemShut
  {NoStop}%
\bibitem [{\citenamefont {Bernard}\ \emph {et~al.}(2006)\citenamefont
  {Bernard}, \citenamefont {Oertel}, \citenamefont {Passemar},\ and\
  \citenamefont {Stern}}]{Bernard:2006gy}%
  \BibitemOpen
  \bibfield  {author} {\bibinfo {author} {\bibfnamefont {V.}~\bibnamefont
  {Bernard}}, \bibinfo {author} {\bibfnamefont {M.}~\bibnamefont {Oertel}},
  \bibinfo {author} {\bibfnamefont {E.}~\bibnamefont {Passemar}}, \ and\
  \bibinfo {author} {\bibfnamefont {J.}~\bibnamefont {Stern}},\ }\href
  {\doibase 10.1016/j.physletb.2006.05.079} {\bibfield  {journal} {\bibinfo
  {journal} {Phys. Lett.}\ }\textbf {\bibinfo {volume} {B638}},\ \bibinfo
  {pages} {480} (\bibinfo {year} {2006})},\ \Eprint
  {http://arxiv.org/abs/hep-ph/0603202} {arXiv:hep-ph/0603202 [hep-ph]}
  \BibitemShut {NoStop}%
\bibitem [{\citenamefont {Gonzalez-Alonso}\ \emph {et~al.}(2017)\citenamefont
  {Gonzalez-Alonso}, \citenamefont {Martin~Camalich},\ and\ \citenamefont
  {Mimouni}}]{Gonzalez-Alonso:2017iyc}%
  \BibitemOpen
  \bibfield  {author} {\bibinfo {author} {\bibfnamefont {M.}~\bibnamefont
  {Gonzalez-Alonso}}, \bibinfo {author} {\bibfnamefont {J.}~\bibnamefont
  {Martin~Camalich}}, \ and\ \bibinfo {author} {\bibfnamefont {K.}~\bibnamefont
  {Mimouni}},\ }\href {\doibase 10.1016/j.physletb.2017.07.003} {\bibfield
  {journal} {\bibinfo  {journal} {Phys. Lett.}\ }\textbf {\bibinfo {volume}
  {B772}},\ \bibinfo {pages} {777} (\bibinfo {year} {2017})},\ \Eprint
  {http://arxiv.org/abs/1706.00410} {arXiv:1706.00410 [hep-ph]} \BibitemShut
  {NoStop}%
\bibitem [{\citenamefont {Alioli}\ \emph {et~al.}(2017)\citenamefont {Alioli},
  \citenamefont {Cirigliano}, \citenamefont {Dekens}, \citenamefont
  {de~Vries},\ and\ \citenamefont {Mereghetti}}]{Alioli:2017ces}%
  \BibitemOpen
  \bibfield  {author} {\bibinfo {author} {\bibfnamefont {S.}~\bibnamefont
  {Alioli}}, \bibinfo {author} {\bibfnamefont {V.}~\bibnamefont {Cirigliano}},
  \bibinfo {author} {\bibfnamefont {W.}~\bibnamefont {Dekens}}, \bibinfo
  {author} {\bibfnamefont {J.}~\bibnamefont {de~Vries}}, \ and\ \bibinfo
  {author} {\bibfnamefont {E.}~\bibnamefont {Mereghetti}},\ }\href {\doibase
  10.1007/JHEP05(2017)086} {\bibfield  {journal} {\bibinfo  {journal} {JHEP}\
  }\textbf {\bibinfo {volume} {05}},\ \bibinfo {pages} {086} (\bibinfo {year}
  {2017})},\ \Eprint {http://arxiv.org/abs/1703.04751} {arXiv:1703.04751
  [hep-ph]} \BibitemShut {NoStop}%
\bibitem [{\citenamefont {Chang}\ \emph {et~al.}(2018)\citenamefont {Chang}
  \emph {et~al.}}]{Chang:2018uxx}%
  \BibitemOpen
  \bibfield  {author} {\bibinfo {author} {\bibfnamefont {C.~C.}\ \bibnamefont
  {Chang}} \emph {et~al.},\ }\href {\doibase 10.1038/s41586-018-0161-8}
  {\bibfield  {journal} {\bibinfo  {journal} {Nature}\ }\textbf {\bibinfo
  {volume} {558}},\ \bibinfo {pages} {91} (\bibinfo {year} {2018})},\ \Eprint
  {http://arxiv.org/abs/1805.12130} {arXiv:1805.12130 [hep-lat]} \BibitemShut
  {NoStop}%
\bibitem [{\citenamefont {Gupta}\ \emph {et~al.}(2018)\citenamefont {Gupta},
  \citenamefont {Jang}, \citenamefont {Yoon}, \citenamefont {Lin},
  \citenamefont {Cirigliano},\ and\ \citenamefont
  {Bhattacharya}}]{Gupta:2018qil}%
  \BibitemOpen
  \bibfield  {author} {\bibinfo {author} {\bibfnamefont {R.}~\bibnamefont
  {Gupta}}, \bibinfo {author} {\bibfnamefont {Y.-C.}\ \bibnamefont {Jang}},
  \bibinfo {author} {\bibfnamefont {B.}~\bibnamefont {Yoon}}, \bibinfo {author}
  {\bibfnamefont {H.-W.}\ \bibnamefont {Lin}}, \bibinfo {author} {\bibfnamefont
  {V.}~\bibnamefont {Cirigliano}}, \ and\ \bibinfo {author} {\bibfnamefont
  {T.}~\bibnamefont {Bhattacharya}},\ }\href {\doibase
  10.1103/PhysRevD.98.034503} {\bibfield  {journal} {\bibinfo  {journal} {Phys.
  Rev.}\ }\textbf {\bibinfo {volume} {D98}},\ \bibinfo {pages} {034503}
  (\bibinfo {year} {2018})},\ \Eprint {http://arxiv.org/abs/1806.09006}
  {arXiv:1806.09006 [hep-lat]} \BibitemShut {NoStop}%
\bibitem [{\citenamefont {Aaboud}\ \emph {et~al.}(2018)\citenamefont {Aaboud}
  \emph {et~al.}}]{Aaboud:2018vgh}%
  \BibitemOpen
  \bibfield  {author} {\bibinfo {author} {\bibfnamefont {M.}~\bibnamefont
  {Aaboud}} \emph {et~al.} (\bibinfo {collaboration} {ATLAS}),\ }\href@noop {}
  {\  (\bibinfo {year} {2018})},\ \Eprint {http://arxiv.org/abs/1801.06992}
  {arXiv:1801.06992 [hep-ex]} \BibitemShut {NoStop}%
\bibitem [{\citenamefont {Alwall}\ \emph {et~al.}(2014)\citenamefont {Alwall},
  \citenamefont {Frederix}, \citenamefont {Frixione}, \citenamefont {Hirschi},
  \citenamefont {Maltoni} \emph {et~al.}}]{Alwall:2014hca}%
  \BibitemOpen
  \bibfield  {author} {\bibinfo {author} {\bibfnamefont {J.}~\bibnamefont
  {Alwall}}, \bibinfo {author} {\bibfnamefont {R.}~\bibnamefont {Frederix}},
  \bibinfo {author} {\bibfnamefont {S.}~\bibnamefont {Frixione}}, \bibinfo
  {author} {\bibfnamefont {V.}~\bibnamefont {Hirschi}}, \bibinfo {author}
  {\bibfnamefont {F.}~\bibnamefont {Maltoni}},  \emph {et~al.},\ }\href
  {\doibase 10.1007/JHEP07(2014)079} {\bibfield  {journal} {\bibinfo  {journal}
  {JHEP}\ }\textbf {\bibinfo {volume} {1407}},\ \bibinfo {pages} {079}
  (\bibinfo {year} {2014})},\ \Eprint {http://arxiv.org/abs/1405.0301}
  {arXiv:1405.0301 [hep-ph]} \BibitemShut {NoStop}%
\bibitem [{\citenamefont {Sjostrand}\ \emph {et~al.}(2015)\citenamefont
  {Sjostrand}, \citenamefont {Ask}, \citenamefont {Christiansen}, \citenamefont
  {Corke}, \citenamefont {Desai}, \citenamefont {Ilten}, \citenamefont
  {Mrenna}, \citenamefont {Prestel}, \citenamefont {Rasmussen},\ and\
  \citenamefont {Skands}}]{Sjostrand:2014zea}%
  \BibitemOpen
  \bibfield  {author} {\bibinfo {author} {\bibfnamefont {T.}~\bibnamefont
  {Sjostrand}}, \bibinfo {author} {\bibfnamefont {S.}~\bibnamefont {Ask}},
  \bibinfo {author} {\bibfnamefont {J.~R.}\ \bibnamefont {Christiansen}},
  \bibinfo {author} {\bibfnamefont {R.}~\bibnamefont {Corke}}, \bibinfo
  {author} {\bibfnamefont {N.}~\bibnamefont {Desai}}, \bibinfo {author}
  {\bibfnamefont {P.}~\bibnamefont {Ilten}}, \bibinfo {author} {\bibfnamefont
  {S.}~\bibnamefont {Mrenna}}, \bibinfo {author} {\bibfnamefont
  {S.}~\bibnamefont {Prestel}}, \bibinfo {author} {\bibfnamefont {C.~O.}\
  \bibnamefont {Rasmussen}}, \ and\ \bibinfo {author} {\bibfnamefont {P.~Z.}\
  \bibnamefont {Skands}},\ }\href {\doibase 10.1016/j.cpc.2015.01.024}
  {\bibfield  {journal} {\bibinfo  {journal} {Comput. Phys. Commun.}\ }\textbf
  {\bibinfo {volume} {191}},\ \bibinfo {pages} {159} (\bibinfo {year}
  {2015})},\ \Eprint {http://arxiv.org/abs/1410.3012} {arXiv:1410.3012
  [hep-ph]} \BibitemShut {NoStop}%
\bibitem [{\citenamefont {de~Favereau}\ \emph {et~al.}(2013)\citenamefont
  {de~Favereau}, \citenamefont {Delaere}, \citenamefont {Demin}, \citenamefont
  {Giammanco}, \citenamefont {Lemaître} \emph {et~al.}}]{deFavereau:2013fsa}%
  \BibitemOpen
  \bibfield  {author} {\bibinfo {author} {\bibfnamefont {J.}~\bibnamefont
  {de~Favereau}}, \bibinfo {author} {\bibfnamefont {C.}~\bibnamefont
  {Delaere}}, \bibinfo {author} {\bibfnamefont {P.}~\bibnamefont {Demin}},
  \bibinfo {author} {\bibfnamefont {A.}~\bibnamefont {Giammanco}}, \bibinfo
  {author} {\bibfnamefont {V.}~\bibnamefont {Lemaître}},  \emph {et~al.},\
  }\href@noop {} {\  (\bibinfo {year} {2013})},\ \Eprint
  {http://arxiv.org/abs/1307.6346} {arXiv:1307.6346 [hep-ex]} \BibitemShut
  {NoStop}%
\bibitem [{\citenamefont {Alioli}\ \emph {et~al.}(2018)\citenamefont {Alioli},
  \citenamefont {Dekens}, \citenamefont {Girard},\ and\ \citenamefont
  {Mereghetti}}]{Alioli:2018ljm}%
  \BibitemOpen
  \bibfield  {author} {\bibinfo {author} {\bibfnamefont {S.}~\bibnamefont
  {Alioli}}, \bibinfo {author} {\bibfnamefont {W.}~\bibnamefont {Dekens}},
  \bibinfo {author} {\bibfnamefont {M.}~\bibnamefont {Girard}}, \ and\ \bibinfo
  {author} {\bibfnamefont {E.}~\bibnamefont {Mereghetti}},\ }\href@noop {} {\
  (\bibinfo {year} {2018})},\ \Eprint {http://arxiv.org/abs/1804.07407}
  {arXiv:1804.07407 [hep-ph]} \BibitemShut {NoStop}%
\bibitem [{\citenamefont {Greljo}\ and\ \citenamefont
  {Marzocca}(2017)}]{Greljo:2017vvb}%
  \BibitemOpen
  \bibfield  {author} {\bibinfo {author} {\bibfnamefont {A.}~\bibnamefont
  {Greljo}}\ and\ \bibinfo {author} {\bibfnamefont {D.}~\bibnamefont
  {Marzocca}},\ }\href {\doibase 10.1140/epjc/s10052-017-5119-8} {\bibfield
  {journal} {\bibinfo  {journal} {Eur. Phys. J.}\ }\textbf {\bibinfo {volume}
  {C77}},\ \bibinfo {pages} {548} (\bibinfo {year} {2017})},\ \Eprint
  {http://arxiv.org/abs/1704.09015} {arXiv:1704.09015 [hep-ph]} \BibitemShut
  {NoStop}%
\bibitem [{\citenamefont {Lees}\ \emph {et~al.}(2013)\citenamefont {Lees} \emph
  {et~al.}}]{Lees:2013uzd}%
  \BibitemOpen
  \bibfield  {author} {\bibinfo {author} {\bibfnamefont {J.~P.}\ \bibnamefont
  {Lees}} \emph {et~al.} (\bibinfo {collaboration} {BaBar}),\ }\href {\doibase
  10.1103/PhysRevD.88.072012} {\bibfield  {journal} {\bibinfo  {journal} {Phys.
  Rev.}\ }\textbf {\bibinfo {volume} {D88}},\ \bibinfo {pages} {072012}
  (\bibinfo {year} {2013})},\ \Eprint {http://arxiv.org/abs/1303.0571}
  {arXiv:1303.0571 [hep-ex]} \BibitemShut {NoStop}%
\bibitem [{\citenamefont {Huschle}\ \emph {et~al.}(2015)\citenamefont {Huschle}
  \emph {et~al.}}]{Huschle:2015rga}%
  \BibitemOpen
  \bibfield  {author} {\bibinfo {author} {\bibfnamefont {M.}~\bibnamefont
  {Huschle}} \emph {et~al.} (\bibinfo {collaboration} {Belle}),\ }\href
  {\doibase 10.1103/PhysRevD.92.072014} {\bibfield  {journal} {\bibinfo
  {journal} {Phys. Rev.}\ }\textbf {\bibinfo {volume} {D92}},\ \bibinfo {pages}
  {072014} (\bibinfo {year} {2015})},\ \Eprint
  {http://arxiv.org/abs/1507.03233} {arXiv:1507.03233 [hep-ex]} \BibitemShut
  {NoStop}%
\bibitem [{\citenamefont {Sato}\ \emph {et~al.}(2016)\citenamefont {Sato} \emph
  {et~al.}}]{Sato:2016svk}%
  \BibitemOpen
  \bibfield  {author} {\bibinfo {author} {\bibfnamefont {Y.}~\bibnamefont
  {Sato}} \emph {et~al.} (\bibinfo {collaboration} {Belle}),\ }\href {\doibase
  10.1103/PhysRevD.94.072007} {\bibfield  {journal} {\bibinfo  {journal} {Phys.
  Rev.}\ }\textbf {\bibinfo {volume} {D94}},\ \bibinfo {pages} {072007}
  (\bibinfo {year} {2016})},\ \Eprint {http://arxiv.org/abs/1607.07923}
  {arXiv:1607.07923 [hep-ex]} \BibitemShut {NoStop}%
\bibitem [{\citenamefont {Aaij}\ \emph {et~al.}(2014)\citenamefont {Aaij} \emph
  {et~al.}}]{Aaij:2014ora}%
  \BibitemOpen
  \bibfield  {author} {\bibinfo {author} {\bibfnamefont {R.}~\bibnamefont
  {Aaij}} \emph {et~al.} (\bibinfo {collaboration} {LHCb}),\ }\href {\doibase
  10.1103/PhysRevLett.113.151601} {\bibfield  {journal} {\bibinfo  {journal}
  {Phys. Rev. Lett.}\ }\textbf {\bibinfo {volume} {113}},\ \bibinfo {pages}
  {151601} (\bibinfo {year} {2014})},\ \Eprint {http://arxiv.org/abs/1406.6482}
  {arXiv:1406.6482 [hep-ex]} \BibitemShut {NoStop}%
\bibitem [{\citenamefont {Aaij}\ \emph {et~al.}(2017)\citenamefont {Aaij} \emph
  {et~al.}}]{Aaij:2017vbb}%
  \BibitemOpen
  \bibfield  {author} {\bibinfo {author} {\bibfnamefont {R.}~\bibnamefont
  {Aaij}} \emph {et~al.} (\bibinfo {collaboration} {LHCb}),\ }\href {\doibase
  10.1007/JHEP08(2017)055} {\bibfield  {journal} {\bibinfo  {journal} {JHEP}\
  }\textbf {\bibinfo {volume} {08}},\ \bibinfo {pages} {055} (\bibinfo {year}
  {2017})},\ \Eprint {http://arxiv.org/abs/1705.05802} {arXiv:1705.05802
  [hep-ex]} \BibitemShut {NoStop}%
\bibitem [{\citenamefont {Filipuzzi}\ \emph {et~al.}(2012)\citenamefont
  {Filipuzzi}, \citenamefont {Portoles},\ and\ \citenamefont
  {Gonzalez-Alonso}}]{Filipuzzi:2012mg}%
  \BibitemOpen
  \bibfield  {author} {\bibinfo {author} {\bibfnamefont {A.}~\bibnamefont
  {Filipuzzi}}, \bibinfo {author} {\bibfnamefont {J.}~\bibnamefont {Portoles}},
  \ and\ \bibinfo {author} {\bibfnamefont {M.}~\bibnamefont
  {Gonzalez-Alonso}},\ }\href {\doibase 10.1103/PhysRevD.85.116010} {\bibfield
  {journal} {\bibinfo  {journal} {Phys. Rev.}\ }\textbf {\bibinfo {volume}
  {D85}},\ \bibinfo {pages} {116010} (\bibinfo {year} {2012})},\ \Eprint
  {http://arxiv.org/abs/1203.2092} {arXiv:1203.2092 [hep-ph]} \BibitemShut
  {NoStop}%
\bibitem [{\citenamefont {Shifman}\ \emph {et~al.}(1979)\citenamefont
  {Shifman}, \citenamefont {Vainshtein},\ and\ \citenamefont
  {Zakharov}}]{Shifman:1978bx}%
  \BibitemOpen
  \bibfield  {author} {\bibinfo {author} {\bibfnamefont {M.~A.}\ \bibnamefont
  {Shifman}}, \bibinfo {author} {\bibfnamefont {A.~I.}\ \bibnamefont
  {Vainshtein}}, \ and\ \bibinfo {author} {\bibfnamefont {V.~I.}\ \bibnamefont
  {Zakharov}},\ }\href {\doibase 10.1016/0550-3213(79)90022-1} {\bibfield
  {journal} {\bibinfo  {journal} {Nucl. Phys.}\ }\textbf {\bibinfo {volume}
  {B147}},\ \bibinfo {pages} {385} (\bibinfo {year} {1979})}\BibitemShut
  {NoStop}%
\bibitem [{\citenamefont {Baikov}\ \emph {et~al.}(2008)\citenamefont {Baikov},
  \citenamefont {Chetyrkin},\ and\ \citenamefont {Kuhn}}]{Baikov:2008jh}%
  \BibitemOpen
  \bibfield  {author} {\bibinfo {author} {\bibfnamefont {P.~A.}\ \bibnamefont
  {Baikov}}, \bibinfo {author} {\bibfnamefont {K.~G.}\ \bibnamefont
  {Chetyrkin}}, \ and\ \bibinfo {author} {\bibfnamefont {J.~H.}\ \bibnamefont
  {Kuhn}},\ }\href {\doibase 10.1103/PhysRevLett.101.012002} {\bibfield
  {journal} {\bibinfo  {journal} {Phys. Rev. Lett.}\ }\textbf {\bibinfo
  {volume} {101}},\ \bibinfo {pages} {012002} (\bibinfo {year} {2008})},\
  \Eprint {http://arxiv.org/abs/0801.1821} {arXiv:0801.1821 [hep-ph]}
  \BibitemShut {NoStop}%
\bibitem [{\citenamefont {Pivovarov}(1992)}]{Pivovarov:1991rh}%
  \BibitemOpen
  \bibfield  {author} {\bibinfo {author} {\bibfnamefont {A.~A.}\ \bibnamefont
  {Pivovarov}},\ }\href {\doibase 10.1007/BF01625906} {\bibfield  {journal}
  {\bibinfo  {journal} {Z. Phys.}\ }\textbf {\bibinfo {volume} {C53}},\
  \bibinfo {pages} {461} (\bibinfo {year} {1992})},\ \bibinfo {note} {[Yad.
  Fiz.54,1114(1991)]},\ \Eprint {http://arxiv.org/abs/hep-ph/0302003}
  {arXiv:hep-ph/0302003 [hep-ph]} \BibitemShut {NoStop}%
\bibitem [{\citenamefont {Le~Diberder}\ and\ \citenamefont
  {Pich}(1992)}]{LeDiberder:1992te}%
  \BibitemOpen
  \bibfield  {author} {\bibinfo {author} {\bibfnamefont {F.}~\bibnamefont
  {Le~Diberder}}\ and\ \bibinfo {author} {\bibfnamefont {A.}~\bibnamefont
  {Pich}},\ }\href {\doibase 10.1016/0370-2693(92)90172-Z} {\bibfield
  {journal} {\bibinfo  {journal} {Phys. Lett.}\ }\textbf {\bibinfo {volume}
  {B286}},\ \bibinfo {pages} {147} (\bibinfo {year} {1992})}\BibitemShut
  {NoStop}%
\bibitem [{\citenamefont {Donoghue}\ and\ \citenamefont
  {Golowich}(2000)}]{Donoghue:1999ku}%
  \BibitemOpen
  \bibfield  {author} {\bibinfo {author} {\bibfnamefont {J.~F.}\ \bibnamefont
  {Donoghue}}\ and\ \bibinfo {author} {\bibfnamefont {E.}~\bibnamefont
  {Golowich}},\ }\href {\doibase 10.1016/S0370-2693(00)00239-2} {\bibfield
  {journal} {\bibinfo  {journal} {Phys. Lett.}\ }\textbf {\bibinfo {volume}
  {B478}},\ \bibinfo {pages} {172} (\bibinfo {year} {2000})},\ \Eprint
  {http://arxiv.org/abs/hep-ph/9911309} {arXiv:hep-ph/9911309} \BibitemShut
  {NoStop}%
\bibitem [{\citenamefont {Cirigliano}\ \emph {et~al.}(2001)\citenamefont
  {Cirigliano}, \citenamefont {Donoghue}, \citenamefont {Golowich},\ and\
  \citenamefont {Maltman}}]{Cirigliano:2001qw}%
  \BibitemOpen
  \bibfield  {author} {\bibinfo {author} {\bibfnamefont {V.}~\bibnamefont
  {Cirigliano}}, \bibinfo {author} {\bibfnamefont {J.~F.}\ \bibnamefont
  {Donoghue}}, \bibinfo {author} {\bibfnamefont {E.}~\bibnamefont {Golowich}},
  \ and\ \bibinfo {author} {\bibfnamefont {K.}~\bibnamefont {Maltman}},\ }\href
  {\doibase 10.1016/S0370-2693(01)01250-3} {\bibfield  {journal} {\bibinfo
  {journal} {Phys. Lett.}\ }\textbf {\bibinfo {volume} {B522}},\ \bibinfo
  {pages} {245} (\bibinfo {year} {2001})},\ \Eprint
  {http://arxiv.org/abs/hep-ph/0109113} {arXiv:hep-ph/0109113} \BibitemShut
  {NoStop}%
\bibitem [{\citenamefont {Cirigliano}\ \emph {et~al.}(2003)\citenamefont
  {Cirigliano}, \citenamefont {Donoghue}, \citenamefont {Golowich},\ and\
  \citenamefont {Maltman}}]{Cirigliano:2002jy}%
  \BibitemOpen
  \bibfield  {author} {\bibinfo {author} {\bibfnamefont {V.}~\bibnamefont
  {Cirigliano}}, \bibinfo {author} {\bibfnamefont {J.~F.}\ \bibnamefont
  {Donoghue}}, \bibinfo {author} {\bibfnamefont {E.}~\bibnamefont {Golowich}},
  \ and\ \bibinfo {author} {\bibfnamefont {K.}~\bibnamefont {Maltman}},\ }\href
  {\doibase 10.1016/S0370-2693(03)00010-8} {\bibfield  {journal} {\bibinfo
  {journal} {Phys. Lett.}\ }\textbf {\bibinfo {volume} {B555}},\ \bibinfo
  {pages} {71} (\bibinfo {year} {2003})},\ \Eprint
  {http://arxiv.org/abs/hep-ph/0211420} {arXiv:hep-ph/0211420} \BibitemShut
  {NoStop}%
\bibitem [{\citenamefont {Blum}\ \emph {et~al.}(2012)\citenamefont {Blum} \emph
  {et~al.}}]{Blum:2012uk}%
  \BibitemOpen
  \bibfield  {author} {\bibinfo {author} {\bibfnamefont {T.}~\bibnamefont
  {Blum}} \emph {et~al.},\ }\href {\doibase 10.1103/PhysRevD.86.074513}
  {\bibfield  {journal} {\bibinfo  {journal} {Phys. Rev.}\ }\textbf {\bibinfo
  {volume} {D86}},\ \bibinfo {pages} {074513} (\bibinfo {year} {2012})},\
  \Eprint {http://arxiv.org/abs/1206.5142} {arXiv:1206.5142 [hep-lat]}
  \BibitemShut {NoStop}%
\bibitem [{\citenamefont {Balitsky}\ \emph {et~al.}(1985)\citenamefont
  {Balitsky}, \citenamefont {Kolesnichenko},\ and\ \citenamefont
  {Yung}}]{Balitsky:1985aq}%
  \BibitemOpen
  \bibfield  {author} {\bibinfo {author} {\bibfnamefont {I.~I.}\ \bibnamefont
  {Balitsky}}, \bibinfo {author} {\bibfnamefont {A.~V.}\ \bibnamefont
  {Kolesnichenko}}, \ and\ \bibinfo {author} {\bibfnamefont {A.~V.}\
  \bibnamefont {Yung}},\ }\href@noop {} {\bibfield  {journal} {\bibinfo
  {journal} {Sov. J. Nucl. Phys.}\ }\textbf {\bibinfo {volume} {41}},\ \bibinfo
  {pages} {178} (\bibinfo {year} {1985})},\ \bibinfo {note} {[Yad.
  Fiz.41,282(1985)]}\BibitemShut {NoStop}%
\bibitem [{\citenamefont {Cossu}\ \emph {et~al.}(2016)\citenamefont {Cossu},
  \citenamefont {Fukaya}, \citenamefont {Hashimoto}, \citenamefont {Kaneko},\
  and\ \citenamefont {Noaki}}]{Cossu:2016eqs}%
  \BibitemOpen
  \bibfield  {author} {\bibinfo {author} {\bibfnamefont {G.}~\bibnamefont
  {Cossu}}, \bibinfo {author} {\bibfnamefont {H.}~\bibnamefont {Fukaya}},
  \bibinfo {author} {\bibfnamefont {S.}~\bibnamefont {Hashimoto}}, \bibinfo
  {author} {\bibfnamefont {T.}~\bibnamefont {Kaneko}}, \ and\ \bibinfo {author}
  {\bibfnamefont {J.-I.}\ \bibnamefont {Noaki}},\ }\href {\doibase
  10.1093/ptep/ptw129} {\bibfield  {journal} {\bibinfo  {journal} {PTEP}\
  }\textbf {\bibinfo {volume} {2016}},\ \bibinfo {pages} {093B06} (\bibinfo
  {year} {2016})},\ \Eprint {http://arxiv.org/abs/1607.01099} {arXiv:1607.01099
  [hep-lat]} \BibitemShut {NoStop}%
\bibitem [{\citenamefont {Durr}\ \emph {et~al.}(2014)\citenamefont {Durr} \emph
  {et~al.}}]{Durr:2013goa}%
  \BibitemOpen
  \bibfield  {author} {\bibinfo {author} {\bibfnamefont {S.}~\bibnamefont
  {Durr}} \emph {et~al.} (\bibinfo {collaboration}
  {Budapest-Marseille-Wuppertal}),\ }\href {\doibase
  10.1103/PhysRevD.90.114504} {\bibfield  {journal} {\bibinfo  {journal} {Phys.
  Rev.}\ }\textbf {\bibinfo {volume} {D90}},\ \bibinfo {pages} {114504}
  (\bibinfo {year} {2014})},\ \Eprint {http://arxiv.org/abs/1310.3626}
  {arXiv:1310.3626 [hep-lat]} \BibitemShut {NoStop}%
\bibitem [{\citenamefont {Bazavov}\ \emph
  {et~al.}(2010{\natexlab{b}})\citenamefont {Bazavov} \emph
  {et~al.}}]{Bazavov:2010yq}%
  \BibitemOpen
  \bibfield  {author} {\bibinfo {author} {\bibfnamefont {A.}~\bibnamefont
  {Bazavov}} \emph {et~al.},\ }\bibfield  {booktitle} {\emph {\bibinfo
  {booktitle} {{Proceedings, 28th International Symposium on Lattice field
  theory (Lattice 2010): Villasimius, Italy, June 14-19, 2010}}},\ }\href@noop
  {} {\bibfield  {journal} {\bibinfo  {journal} {PoS}\ }\textbf {\bibinfo
  {volume} {LATTICE2010}},\ \bibinfo {pages} {083} (\bibinfo {year}
  {2010}{\natexlab{b}})},\ \Eprint {http://arxiv.org/abs/1011.1792}
  {arXiv:1011.1792 [hep-lat]} \BibitemShut {NoStop}%
\bibitem [{\citenamefont {Borsanyi}\ \emph {et~al.}(2013)\citenamefont
  {Borsanyi}, \citenamefont {Durr}, \citenamefont {Fodor}, \citenamefont
  {Krieg}, \citenamefont {Schafer}, \citenamefont {Scholz},\ and\ \citenamefont
  {Szabo}}]{Borsanyi:2012zv}%
  \BibitemOpen
  \bibfield  {author} {\bibinfo {author} {\bibfnamefont {S.}~\bibnamefont
  {Borsanyi}}, \bibinfo {author} {\bibfnamefont {S.}~\bibnamefont {Durr}},
  \bibinfo {author} {\bibfnamefont {Z.}~\bibnamefont {Fodor}}, \bibinfo
  {author} {\bibfnamefont {S.}~\bibnamefont {Krieg}}, \bibinfo {author}
  {\bibfnamefont {A.}~\bibnamefont {Schafer}}, \bibinfo {author} {\bibfnamefont
  {E.~E.}\ \bibnamefont {Scholz}}, \ and\ \bibinfo {author} {\bibfnamefont
  {K.~K.}\ \bibnamefont {Szabo}},\ }\href {\doibase 10.1103/PhysRevD.88.014513}
  {\bibfield  {journal} {\bibinfo  {journal} {Phys. Rev.}\ }\textbf {\bibinfo
  {volume} {D88}},\ \bibinfo {pages} {014513} (\bibinfo {year} {2013})},\
  \Eprint {http://arxiv.org/abs/1205.0788} {arXiv:1205.0788 [hep-lat]}
  \BibitemShut {NoStop}%
\bibitem [{\citenamefont {Boyle}\ \emph {et~al.}(2016)\citenamefont {Boyle}
  \emph {et~al.}}]{Boyle:2015exm}%
  \BibitemOpen
  \bibfield  {author} {\bibinfo {author} {\bibfnamefont {P.~A.}\ \bibnamefont
  {Boyle}} \emph {et~al.},\ }\href {\doibase 10.1103/PhysRevD.93.054502}
  {\bibfield  {journal} {\bibinfo  {journal} {Phys. Rev.}\ }\textbf {\bibinfo
  {volume} {D93}},\ \bibinfo {pages} {054502} (\bibinfo {year} {2016})},\
  \Eprint {http://arxiv.org/abs/1511.01950} {arXiv:1511.01950 [hep-lat]}
  \BibitemShut {NoStop}%
\bibitem [{\citenamefont {Chibisov}\ \emph {et~al.}(1997)\citenamefont
  {Chibisov}, \citenamefont {Dikeman}, \citenamefont {Shifman},\ and\
  \citenamefont {Uraltsev}}]{Chibisov:1996wf}%
  \BibitemOpen
  \bibfield  {author} {\bibinfo {author} {\bibfnamefont {B.}~\bibnamefont
  {Chibisov}}, \bibinfo {author} {\bibfnamefont {R.~D.}\ \bibnamefont
  {Dikeman}}, \bibinfo {author} {\bibfnamefont {M.~A.}\ \bibnamefont
  {Shifman}}, \ and\ \bibinfo {author} {\bibfnamefont {N.}~\bibnamefont
  {Uraltsev}},\ }\href {\doibase 10.1142/S0217751X97001316} {\bibfield
  {journal} {\bibinfo  {journal} {Int. J. Mod. Phys.}\ }\textbf {\bibinfo
  {volume} {A12}},\ \bibinfo {pages} {2075} (\bibinfo {year} {1997})},\ \Eprint
  {http://arxiv.org/abs/hep-ph/9605465} {arXiv:hep-ph/9605465 [hep-ph]}
  \BibitemShut {NoStop}%
\bibitem [{\citenamefont {Shifman}(2001)}]{Shifman:2000jv}%
  \BibitemOpen
  \bibfield  {author} {\bibinfo {author} {\bibfnamefont {M.~A.}\ \bibnamefont
  {Shifman}},\ }in\ \href {\doibase 10.1142/9789812810458_0032} {\emph
  {\bibinfo {booktitle} {{At the frontier of particle physics. Handbook of QCD.
  Vol. 1-3}}}},\ \bibinfo {organization} {World Scientific}\ (\bibinfo
  {publisher} {World Scientific},\ \bibinfo {address} {Singapore},\ \bibinfo
  {year} {2001})\ pp.\ \bibinfo {pages} {1447--1494},\ \bibinfo {note}
  {[3,1447(2000)]},\ \Eprint {http://arxiv.org/abs/hep-ph/0009131}
  {arXiv:hep-ph/0009131 [hep-ph]} \BibitemShut {NoStop}%
\bibitem [{\citenamefont {Cata}\ \emph {et~al.}(2005)\citenamefont {Cata},
  \citenamefont {Golterman},\ and\ \citenamefont {Peris}}]{Cata:2005zj}%
  \BibitemOpen
  \bibfield  {author} {\bibinfo {author} {\bibfnamefont {O.}~\bibnamefont
  {Cata}}, \bibinfo {author} {\bibfnamefont {M.}~\bibnamefont {Golterman}}, \
  and\ \bibinfo {author} {\bibfnamefont {S.}~\bibnamefont {Peris}},\ }\href
  {\doibase 10.1088/1126-6708/2005/08/076} {\bibfield  {journal} {\bibinfo
  {journal} {JHEP}\ }\textbf {\bibinfo {volume} {08}},\ \bibinfo {pages} {076}
  (\bibinfo {year} {2005})},\ \Eprint {http://arxiv.org/abs/hep-ph/0506004}
  {arXiv:hep-ph/0506004 [hep-ph]} \BibitemShut {NoStop}%
\bibitem [{\citenamefont {Gonzalez-Alonso}\ \emph {et~al.}(2010)\citenamefont
  {Gonzalez-Alonso}, \citenamefont {Pich},\ and\ \citenamefont
  {Prades}}]{GonzalezAlonso:2010xf}%
  \BibitemOpen
  \bibfield  {author} {\bibinfo {author} {\bibfnamefont {M.}~\bibnamefont
  {Gonzalez-Alonso}}, \bibinfo {author} {\bibfnamefont {A.}~\bibnamefont
  {Pich}}, \ and\ \bibinfo {author} {\bibfnamefont {J.}~\bibnamefont
  {Prades}},\ }\href {\doibase 10.1103/PhysRevD.82.014019} {\bibfield
  {journal} {\bibinfo  {journal} {Phys. Rev.}\ }\textbf {\bibinfo {volume}
  {D82}},\ \bibinfo {pages} {014019} (\bibinfo {year} {2010})},\ \Eprint
  {http://arxiv.org/abs/1004.4987} {arXiv:1004.4987 [hep-ph]} \BibitemShut
  {NoStop}%
\end{thebibliography}%

\end{document}